\documentclass[aps, prd, onecolumn, groupedaddress, eqsecnum ]{revtex4-1}
\usepackage{amsmath,amssymb}
\usepackage{graphicx}
\usepackage{color}
\usepackage{slashed} 
\usepackage{hyperref}
\begin{document}
\newcommand{\nn}{\nonumber}
\def\d{{\mathrm{d}}}
\def\lint{\hbox{\Large $\displaystyle\int$}}   
\def\hint{\hbox{\huge $\displaystyle\int$}}  
\title{\bf\Large Simplifying the Reinsch algorithm for the  \\Baker--Campbell--Hausdorff series}
\author{Alexander Van--Brunt\,}
\email[]{alexandervanbrunt@gmail.com}
\author{Matt Visser\,}
\email[]{matt.visser@msor.vuw.ac.nz}
\affiliation{ \mbox{School of Mathematics and Statistics,}
Victoria University of Wellington; \\
PO Box 600, Wellington 6140, New Zealand.\\}
\date{21 January 2015;  27 January 2015; 23 July 2015; 3 December 2015; \LaTeX-ed \today}
\begin{abstract}

The Goldberg version of the Baker--Campbell--Hausdorff series computes the quantity 
\begin{equation*}
Z(X,Y)=\ln\left( e^X e^Y \right) = \sum_w g(w) \;  w(X,Y)
\end{equation*}
where $X$ and $Y$ are not necessarily commuting, in terms of ``words'' constructed from the $\{X,Y\}$ ``alphabet''. The so-called Goldberg coefficients $g(w)$ are the  central topic of this article. 
This Baker--Campbell--Hausdorff series is a general purpose tool of very wide applicability in mathematical physics, quantum physics, and many other fields. 
The Reinsch algorithm for the truncated series permits one to calculate the Goldberg coefficients up to some fixed word length $|w|$ by using nilpotent $(|w|+1)\times(|w|+1)$ matrices. 
We shall show how to further simplify the Reinsch algorithm, making its implementation (in principle) utterly straightforward using ``off the shelf'' symbolic manipulation software. 
Specific computations provide examples which help to provide a deeper understanding of the Goldberg coefficients and their properties. 
For instance we shall establish some strict bounds (and some equalities) on the number of non-zero Goldberg coefficients. 
Unfortunately, we shall see that the number of nonzero Goldberg coefficients often grows very rapidly (in fact exponentially) with the word length $|w|$.

Furthermore the simplified Reinsch algorithm readily generalizes to many closely related but still quite distinct problems --- we shall also present closely related results for the symmetric product
\begin{equation*}
S(X,Y)=\ln\left( e^{X/2} e^Y e^{X/2} \right)   = \sum_w g_{\scriptscriptstyle{S}}(w) \;  w(X,Y).
\end{equation*}
Variations on such themes are straightforward. For instance, one can just as easily consider the ``loop'' product
\begin{equation*}
L(X,Y)=\ln\left( e^{X} e^Y e^{-X} e^{-Y}\right) = \sum_w g_{\scriptscriptstyle{L}}(w) \;  w(X,Y). 
\end{equation*}
This ``loop'' type of series is of interest, for instance, when considering either differential geometric parallel transport around a closed curve, non-Abelian versions of Stokes' theorem, or even Wigner rotation/Thomas precession in special relativity. 
Several other closely related series are also briefly investigated.

\null\qquad\qquad\qquad\! Baker--Campbell--Hausdorff formula.

\bigskip
\noindent
{\sc{arXiv}}:  arXiv:1501.05034 [math-ph] 

\bigskip
\noindent
{\sc{Published as}}: Journal of Mathematical Physics {\bf57} (2016) 023507; doi: 10.1063/1.4939929

\end{abstract}
\maketitle
\hrule
\tableofcontents
\bigskip
\hrule
\clearpage
\section{Introduction}
\parskip10pt 
What is now called the  Baker--Campbell--Hausdorff \emph{formula} (\emph{series}) has been studied for well over a century. A recent study of the early history can be found in~\cite{early-history}, with key historical references including~\cite{Baker, Campbell, Hausdorff, Dynkin1, Dynkin2}. A more recent summary of the present-day status of this topic can be found in~\cite{recent-history}. In the current article we will discuss the relatively recently developed Reinsch algorithm~\cite{Reinsch}, and relations between this algorithm and  Goldberg's results from 1956~\cite{Goldberg, Newman-Thompson, Cyclic}.  
Though our primary focus will be the Goldberg coefficients for the non-commutator version of the Baker--Campbell--Hausdorff series, we shall for completeness carefully connect our results back to known results for the nested-commutator versions of the Baker--Campbell--Hausdorff series, and shall furthermore also apply the simplified Reinsch algorithm to several modified variants of the Baker--Campbell--Hausdorff series. Indeed, the simplified Reinsch algorithm can easily be applied to \emph{any} function $f(X,Y)$, of non-commuting variables $X$ and $Y$, for which it is somehow known that a formal power series can be constructed.

One version of the Baker--Campbell--Hausdorff formula is the Goldberg expansion~\cite{Goldberg, Newman-Thompson, Cyclic}
\begin{equation}
Z(X,Y)=\ln( e^X e^Y ) =  \sum_w g(w) \;  w,
\end{equation}
in terms of the rational-number Goldberg coefficients $g(w)$, and ``words'' $w$ constructed from the ``alphabet'' $\{X,Y\}$. 
Letting $|w|$ denote the length of the word $w$, the homogeneous multinomials $z_n(X,Y)$ of degree $n$ are then constructed in terms of words of specified length $n$:
\begin{equation}
z_n(X,Y) = \sum_{|w|=n} g(w)\; w(X,Y).
\end{equation}
It is the efficient computation of these Goldberg coefficients, and some of their analytic properties,  that will be the main focus of this article. 
In counterpoint,  what is now called the  Baker--Campbell--Hausdorff \emph{theorem} is the result that the multinomials $z_n(X,Y)$ are in fact representable in terms of nested commutators (Lie brackets). Specifically, Dynkin's expansion~\cite{Dynkin1, Dynkin2} amounts (with hindsight) to the observation that~\cite{Newman-Thompson, Cyclic}
\begin{equation}
z_n(X,Y) = {1\over n} \sum_{|w|=n} g(w)\; \left[w(X,Y)\right],
\end{equation}
where $\left[w(X,Y)\right]$ denotes the nested commutator built from the word $w(X,Y)$.
For definiteness, we consider right-nested commutators of the form  
\begin{equation}
[ABCDEF\dots] = [A,[B,[C,[D,[E,[F, \dots ]]]]]].
\end{equation}
This is sometimes called the ``long commutator''. (There are also other ways of representing  the commutator terms, see for instance~\cite{Casas-Murua, Murua-tables, Blanes-Casas}, but  for our purposes the right-nested commutators are sufficient.)
Note that, in view of various symmetries, (arising from the antisymmetry of the commutator,  the Jacobi identity, and higher-order commutators inherited from the Jacobi identity), many \emph{different} words $w(X,Y)$ can map to the \emph{same} nested commutator $\left[w(X,Y)\right]$.
Detecting word equivalencies of the type $\left[w_1(X,Y)\right]=\left[w_2(X,Y)\right]$ is a decidedly nontrivial task, which makes the conversion from the Goldberg presentation to one in terms of right-nested commutators computationally expensive. 
Overall we are interested in 
\begin{equation}
Z(X,Y)=\ln\left( e^X e^Y \right) = \sum_w g(w) \;  w =   \sum_{n=1}^\infty \left\{ \sum_{|w|=n} g(w)\; w(X,Y) \right\} =   \sum_{n=1}^\infty z_n(X,Y), 
\end{equation}
and the goal is to find efficient ways of calculating the $g(w)$ and/or the $z_n(X,Y)$. 

Since $\left( e^X e^Y \right)^{-1}= e^{-Y} e^{-X}$ we have $Z(X,Y)=-Z(-Y,-X)$. This implies the two well-known symmetries 
\begin{equation}
z_{2n}(X,Y) = -z_{2n}(Y,X) \qquad\hbox{and} \qquad z_{2n+1}(X,Y) = z_{2n+1}(Y,X). 
\end{equation}
The first few terms in the expansion (converted to nested commutator form) can be explicitly represented as
\begin{eqnarray}
Z(X,Y)&=&\ln( e^X e^Y ) = X + Y + {1\over2}\;[X,Y] + {1\over12}  [X-Y,[X,Y]]  - {1\over24} [X,[Y,[X,Y]]] + \dots
\end{eqnarray}
Unfortunately the expansion rapidly becomes extremely unwieldy.  
Specifically, we shall explicitly prove that  the limit superior of the number of terms in the Goldberg expansion grows  exponentially in $n$. Thus even though explicit computer-aided computations can on a modern laptop easily be carried out to $n=13$, (or sometimes higher if one focusses on specific questions), beyond $n=8$ or thereabouts the resulting formulae are simply too cumbersome to be usefully written down on paper. With a workstation, Casas and Murua~\cite{Casas-Murua} report related calculations out to $n=20$. 
(This style of approach complements what what can be extracted by considering special-case commutators, as for instance in references~\cite{Van-Brunt:special, Van-Brunt:explicit}.)

\section{Simplified algorithm: Low-order terms --- $z_1$ to $z_4$} 
\label{S:low-order}

Our simplified version of Reinsch's algorithm is this: Suppose one wishes to calculate up to some fixed word length $N$. We construct two $(N+1)\times(N+1)$ matrices that are zero except for the first super-diagonal where they contain $N$ distinct elements. That is:
\begin{equation}
X_N =  
\left[ \begin{array}{ccccccc}  
0&x_1&0&0&\cdots &0 &0 \\
0&0&x_2&0&\cdots &0 &0     \\
0&0&0&x_3&\cdots &0 &0     \\
\vdots &\vdots&\vdots& \vdots & \ddots &\vdots& \vdots\\
0&0&0&0&\cdots & x_{N-1}&0     \\
0&0&0&0&\cdots &0& x_N     \\
0&0&0&0&\cdots & 0  &0
\end{array}\right],
\qquad
\hbox{and}
\qquad
Y_N = 
\left[ \begin{array}{ccccccc}  
0&y_1&0&0&\cdots &0 &0 \\
0&0&y_2&0&\cdots &0 &0     \\
0&0&0&y_3&\cdots &0 &0     \\
\vdots &\vdots&\vdots& \vdots & \ddots &\vdots& \vdots\\
0&0&0&0&\cdots & y_{N-1}&0     \\
0&0&0&0&\cdots &0& y_N     \\
0&0&0&0&\cdots & 0  &0
\end{array}\right].
\end{equation}
Now compute, (eg, using {\sf Maple} or some equivalent symbolic algebra software package), the $(N+1)\times(N+1)$ matrix
\begin{equation}
Z_N = \ln\left( \exp(X_N) \exp(Y_N) \right).
\end{equation}
The first row of the matrix $Z_N$ is (essentially) the information we want. Specifically, noting that the matrix $Z_N$ is strictly upper triangular, let us denote
\begin{equation}
Z_N = 
\left[ \begin{array}{ccccccc}  
0&z_1&z_2&z_3&\cdots &z_{N-1} &z_N \\
0&0&*&*&\cdots &* &*     \\
0&0&0&*&\cdots &* &*     \\
\vdots &\vdots&\vdots& \vdots & \ddots &\vdots& \vdots\\
0&0&0&0&\cdots & *&*    \\
0&0&0&0&\cdots  &0& *     \\
0&0&0&0&\cdots & 0  &0
\end{array}\right].
\end{equation}
Here the $*$'s denote nonzero quantities that are not of specific interest for current purposes. 
(The $[Z_N]_{m,m+n}$ are merely copies of the $z_n$, with $n\in(1,N-m)$, and with restricted subscripts $i\in(m,N)$ on the symbols $x_i$ and $y_i$.)
Brute force computation, (see the Appendix for  appropriate {\sf Maple} code), yields as the first four terms:
\begin{eqnarray}
z_1 &=& x_1 + y_1;
\\
z_2 &=& {1\over2}(x_1 y_2- y_1 x_2);
\\
z_3 &=& {1\over12}(x_1 x_2 y_3 -2 x_1 y_2 x_3 +x_1 y_2 y_3 +y_1x_2 x_3-2 y_1 x_2 y_3 +y_1 y_2 x_3);
\\
z_4 &=&{1\over24} (  x_1x_2y_3y_4-2 x_1 y_2x_3y_4+2y_1x_2y_3x_4 - y_1y_2x_3 x_4  ).
\end{eqnarray}
Already at the stage $n=5$ the terms are relatively turgid to explicitly write down. (Full explicit formulae for $z_5$, $z_6$, $z_7$, and $z_8$ are presented in the supplementary material~\cite{supplementary}.)
Next, the corresponding $z_n(X,Y)$ multinomials are constructed by simply replacing $x_i\to X$ and $y_i \to Y$ \emph{while preserving the order of the letters}. That is, for the first four terms,
\begin{eqnarray}
z_1(X,Y) &=& X+Y;\\
z_2(X,Y) &=& {1\over2}(XY-YX);\\
z_3(X,Y) &=& {1\over12}(X^2 Y - 2 XYX + XY^2 +YX^2-2YXY+Y^2X);\\
z_4(X,Y) &=&{1\over24} ( X^2 Y^2 - 2 XYXY +2YXYX - Y^2X^2 ).
\end{eqnarray}
In contrast, the original Reinsch algorithm~\cite{Reinsch} involved an extra set of $N$ formal symbols $\sigma_i$, and an additional  ``symbol conversion stage'', which the current algorithm avoids.  Specifically, Reinsch chose to use
\begin{equation}
F_N =  
\exp \left[ \begin{array}{ccccccc}  
0&1&0&0&\cdots &0 &0 \\
0&0&1&0&\cdots &0 &0     \\
0&0&0&1&\cdots &0 &0     \\
\vdots &\vdots&\vdots& \vdots & \ddots &\vdots& \vdots\\
0&0&0&0&\cdots & 1&0     \\
0&0&0&0&\cdots &0& 1    \\
0&0&0&0&\cdots & 0  &0
\end{array}\right],
\qquad
\hbox{and}
\qquad
G_N = 
\exp \left[ \begin{array}{ccccccc}  
0&\sigma_1&0&0&\cdots &0 &0 \\
0&0&\sigma_2&0&\cdots &0 &0     \\
0&0&0&\sigma_3&\cdots &0 &0     \\
\vdots &\vdots&\vdots& \vdots & \ddots &\vdots& \vdots\\
0&0&0&0&\cdots & \sigma_{N-1}&0     \\
0&0&0&0&\cdots &0& \sigma_N     \\
0&0&0&0&\cdots & 0  &0
\end{array}\right],
\end{equation}
and to then set
\begin{equation}
z_n(X,Y) = T\{\ln( F_N \; G_N)\}_{1,n+1},
\end{equation}
where the $T$-process is a ``translation process'' that maps products of the $\sigma_i$ into ``words'' in the alphabet $\{X,Y\}$ according to the recursive scheme
\begin{equation}
\prod_{i=1}^N  (\sigma_i)^{\mu_i} \to\prod_{i=1}^{N-1}  (\sigma_i)^{\mu_i}  \cup  \{ X \hbox{ if } \mu_N = 0; Y \hbox{ if }  \mu_N = 1  \}.
\end{equation}
One first has to establish that all the $\mu_i \in \{0,1\}$, see~\cite{Reinsch}, and also to check that the $T$-process is a vector-space isomorphism from the space of polynomials in the $\sigma_i$ variables, (with $\mu_i\in\{0,1\}$), to the space of linear combinations of products that have $n$ factors that are either $X$ or $Y$, see~\cite{Reinsch}.
It is this tedious and tricky symbol conversion stage that is completely eliminated in our simplified algorithm,  with all the relevant work being automatically carried out by having the symbolic algebra software simply calculate the quantity $Z_N = \ln\left( \exp(X_N) \exp(Y_N) \right)$. 


The fundamental reason the simplified algorithm works is that $X_N$ and $Y_N$ have been carefully constructed to \emph{not commute} with each other, to be \emph{linearly independent} of each other, and to be \emph{nilpotent}. For instance, for any $m\in\{1,\dots,N\}$ all of the $(X_N)^m$ and $(Y_N)^m$ are non-zero only on the $m$'th super-diagonal, and so these matrices are all linearly independent of each other. Indeed $(X_N)^{N+1}=0=(Y_N)^{N+1}$, establishing nilpotency,  so all matrix functions (in particular the matrix exponential and matrix logarithm) have \emph{finite} Taylor series expansions.  

Specifically, let us consider the closely related $\infty \times \infty$ ``ladder'' matrices $\delta_{i+1,j} \,x $ and $\delta_{i+1,j}\, y$, where the formal symbols $x$ and $y$ need not commute, and the matrices are nonzero only on the first super-diagonal. These matrices possess the property that if we consider any homogeneous polynomial, $P$, of degree $m$, then
$P(\delta_{i+1,j}\, x , \delta_{i+1,j} \,y) = \delta_{i+m, j}\, P(x,y)$, which is non-zero only on the $m^{th}$ super-diagonal. Consequently when we compute the Baker--Campbell--Hausdorff series for these specific matrices, then $\forall i$ the $(i,i+N)$ entry will be exactly all terms of degree $N$ of the Baker--Campbell--Hausdorff  expansion, for arbitrary variables $x$, $y$. In particular, now truncating to $(N+1)\times(N+1)$ matrices,  this truncation naturally eliminates all terms of order greater than $N$; the other non-zero entries in the resulting matrix are the terms of degree less than $N$.
The introduction of the subscripts $x_{i}$ and $y_{i}$ to decorate the elements of these matrices is merely a way of getting around the software's implicit assumption that the variables $x$ and $y$ commute. The elements of the first row of $Z_N$, the $z_n = [Z_N]_{1,n+1}$, are successively built up from a sum of exactly $n$ products of the $X_N$ and $Y_N$. Furthermore $z_n$ will contain a $x_i$ (respectively, a $y_i$) \emph{if and only if} the corresponding string of $n$ matrices has a $X_N$ (respectively, a $Y_N$) in its $i$th position. 
This completes the description of the simplified Reinsch algorithm. 


Subsequently replacing the words $w$  by commutators $[w]$, (for our current purposes an optional but sometimes interesting step), involves several elementary but tricky (and computationally expensive) steps. 
That is, while the simplified Reinsch algorithm is efficient at calculating Goldberg coefficients $g(w)$, this has little impact on efficiently calculating nested commutators.  
To see some of the difficulties involved, note that, from the definition of the long commutator 
\begin{equation}
[ABCDEF\dots] = [A,[B,[C,[D,[E,[F, \dots ]]]]]]
\end{equation}
we immediately have $[wX^{a}] = 0 = [w Y^{a}]$ whenever $a>1$. 
More subtly $[w_1w_2] = [w_1[w_2]]$, which leads to
\begin{equation}
[w_1XYw_2]=[w_1XY[w_2]] = [w_1X[Y,[w_2]]]= [w_1[X,[Y,[w_2]]]].
\end{equation}
Applying the Jacobi identity to this last term yields
\begin{equation}
[w_1XYw_2]=- [w_1[Y,[[w_2],X]]] - [w_1[[w_2],[X,Y]]] = [w_1[Y,[X,[w_2]]]] + [w_1[[X,Y],[w_2]]].
\end{equation}
That is
\begin{equation}
[w_1XYw_2]= [w_1YXw_2] + [w_1[X,Y] w_2].
\end{equation}
This quite general result immediately implies, for instance, identities such as $[YXXY]=[XYXY]$.\\
(Note that $[YXXY]=[Y,[X,[X,Y]]]= -[X,[[X,Y],Y]]-[[X,Y],[Y,X]] =  [X,[Y, [X,Y]]]+[[X,Y],[X,Y]] =  [X,[Y, [X,Y]]] = [XYXY]$.)

Then for instance, when computing the commutator version of  $z_3$, we encounter
\begin{eqnarray}
[X^2 Y - 2 XYX + XY^2 +YX^2-2YXY+Y^2X] &=& [X^2 Y - 2 XYX -2YXY+Y^2X] \nonumber\\
&=& [X^2 Y + 2 X^2Y +2Y^2Y+Y^2X] \nonumber\\
&=& 3 [X^2Y+Y^2X] \nonumber\\
&=& 3 [(X-Y)XY].
\end{eqnarray}
Similarly, when computing the commutator version of $z_4$  one encounters
\begin{eqnarray}
[ X^2 Y^2 - 2 XYXY+2YXYX - Y^2X^2 ] &=& [- 2 XYXY+2YXYX  ]
=- 4 [XYXY].
\end{eqnarray}
This now yields
\begin{eqnarray}
z_1(X,Y) &=& X+Y;\\
z_2(X,Y) &=& {1\over2}[XY];\\
z_3(X,Y) &=& {1\over12} [(X-Y)XY];\\
z_4(X,Y) &=&{1\over24} [XYXY].\\
\end{eqnarray}
For completeness, and more importantly to illustrate some general features of the expansion when carried out to medium-low order, the word versions of $z_5$, $z_6$,  $z_7$, and $z_8$, and the commutator versions of $z_5$ and $z_6$, are presented in the supplementary material \cite{supplementary}. 
The  explicit results for $z_5$, $z_6$,  $z_7$, and $z_8$ are still reasonably tractable.  Beyond this stage the formulae are simply too cumbersome to be usefully written down on paper. The only scientific justification for explicitly presenting even this level of detail is that it explicitly demonstrates the patterns and symmetries of the Goldberg coefficients $g(w)$ in a somewhat non-trivial context. These patterns and symmetries will be useful both in the subsequent Section \ref{S:properties},  in which we investigate some general properties of the Goldberg coefficients, and in the subsequent  Section \ref{S:bounds}, in which we develop bounds on the number of the non-zero Goldberg coefficients.
One reason for going out as far as $z_8$, \emph{and no further},  is that the expression for $z_8$ is actually slightly shorter (by 2 terms) than $z_7$; while in contrast $z_9$ is significantly longer, (slightly over 3 times longer), and can no longer fit on a single sheet of paper.  Another reason for going out to $z_8$, \emph{and no further}, is that $z_8$ contains the first Goldberg coefficient that is intrinsically rational, \begin{equation}
g(X^4Y^4)=g(Y^4X^4) = {23\over120960};
\end{equation}
all other Goldberg coefficients out to 8$^{th}$ order are either zero or reciprocals of integers.


\section{Some properties of the Goldberg coefficients } 
\label{S:properties}

The Goldberg coefficients satisfy a number of interesting analytic and combinatorial properties. (See for instance references~\cite{Goldberg, Newman-Thompson, Cyclic}, and some new results presented below.)
These general properties can easily be checked against the various explicit terms presented in Section~\ref{S:low-order} above and the supplementary material~\cite{supplementary}. 
Some key analytic results are:
\begin{itemize}
\item 
For $n\geq 2$ the sum over words of fixed length is zero:
\begin{equation}
\sum_{|w|=n} g(w) = 0.
\end{equation}
This easily follows from $Z(X,X)=2X$, since the quantity $\sum_{|w|=n} g(w)$ is the coefficient of $X^n$ in $Z(X,X)$, and by inspection this coefficient is seen to be zero. That this sum is zero is a \emph{necessary} condition for the $z_n(X,Y)$ to be representable in terms of nested commutators.

\item 
Let $|w|_X = n_X$ denote the number of letters $X$ in the word $w$, and similarly for $|w|_Y = n_Y$. Then we have $|w|=n=n_X+n_Y$. For $n\geq 2$ the sum over words of fixed $n_X$ and $n_Y$ is zero:
\begin{equation}
\sum_{|w|_X=n_X; \; |w|_Y=n_Y} g(w) = 0.
\end{equation}
This easily follows from $Z(sX,tX)=(s+t)X$, since the quantity $\sum_{|w|_X=n_X; \;|w|_Y=n_Y} g(w)$ is the coefficient of $s^{n_X} t^{n_Y} X^n$ in $Z(sX,tX)$, and by inspection this coefficient is seen to be zero.
In particular this implies $g(X^n)=0=g(Y^n)$, though there are many other ways of convincing oneself of this.

\item
Let $L_i$ denote either of the letters $\{X,Y\}$, so that an arbitrary word $w$ of length $n$ can be represented as 
\begin{equation}
w(L_i,m_i) = L_1^{m_1}L_2^{m_2}\dots L_q^{m_q},
\end{equation}
with $L_i\neq L_{i+1}$ and $m_i\geq 1$, while $n=\sum_{i=1}^q m_i \geq q$. Then~\cite{Goldberg}
\begin{equation}
g(w(L_i,m_i)) = g(w(L_i,m_{\pi(i)})),
\end{equation}
where $\pi(i)$ is an arbitrary permutation of $i\in\{1,2,\dots,q\}$. 
For example $g(X^a Y^b) = g(X^b Y^a)$, 
while $g(X^a Y^b X^c) = g(X^b Y^c X^a) = g(X^c Y^a Y^b) = g(X^c Y^b X^a) = g(X^b Y^a X^c) = g(X^a Y^c X^b)$. 
Identical results hold under the interchange $X \longleftrightarrow Y$. Fundamentally, it is because of this symmetry that so many of the Goldberg coefficients are repeated multiple times. (See Section \ref{S:low-order} above and the supplementary material~\cite{supplementary}.)

\item
Let $L_i$ denote either of the letters $\{X,Y\}$, so that an arbitrary word of length $n$ is $L_1L_2\dots L_n$. (We shall now allow $L_i$ and $L_{i+1}$ to possibly be equal.) Define the cyclic shift operator ${\mathcal C}$ by
\begin{equation}
{\mathcal C}(L_1L_2\dots L_n) = L_2\dots L_n L_1.
\end{equation}
Then acting on words of length $n\geq 2$ we have~\cite{Cyclic}
\begin{equation}
\sum_{m=0}^n  g\left( {\mathcal C}^m (L_1L_2\dots L_n)\right) = 0.
\end{equation}
In particular this implies $g(X^n)=0=g(Y^n)$, though there are many other ways of convincing oneself of this.

\item
Define the  interchange ${\mathcal I}(w)$ of the word $w$ by interchanging all $X\leftrightarrow Y$. 
Then because $Z(X,Y)=-Z(-Y,-X)$ we have
\begin{equation}
g\left( {\mathcal I}(w) \right) = (-1)^{|w|+1} g(w).
\end{equation}

\item
Define the reverse ${\mathcal R}(w)$ of the word $w$ by
\begin{equation}
{\mathcal R}(L_1L_2\dots L_{n-1} L_n) = L_n L_{n-1} \dots L_2 L_1.
\end{equation}
Then
\begin{equation}
g\left( {\mathcal R}\left( {\mathcal I}\left( w\right)\right)\right) = g(w).
\end{equation}
That is, reversing the word and interchanging $X\longleftrightarrow Y$ leaves the Goldberg coefficient invariant. This can be proved by inspection of the explicit formula for $g(w)$ appearing in Section \ref{S:bounds} below. See particularly equation (\ref{E:explicit}) and note that 
\begin{equation}
{\mathcal R}\left( {\mathcal I}\left( X^{r_1} Y^{s_1} X^{r_2} Y^{s_2}  \dots X^{r_k} Y^{s_k}\right)\right)  =X^{s_k} Y^{r_k} X^{s_{k-1}} Y^{r_{k-1}}  \dots X^{s_1} Y^{r_1}.
\end{equation}
Once summed over the $r_i$ and $s_i$ the claimed result follows. 
In particular this implies
\begin{equation}
g\left( {\mathcal R}(w) \right) = (-1)^{|w|+1} \; g(w); \qquad \hbox{and} \qquad g({\mathcal I}(w))=g({\mathcal R}(w)).
\end{equation}
Even more specifically: We see that even length palindromes have Goldberg coefficient zero. For example $g(X^aY^{2b}X^a)=0$. More generally,  concatenating any word $w$ with its reverse $ {\mathcal R}(w)$ one has $g(w\, {\mathcal R}(w))=0$. 

\item
Combining Goldberg's permutation result with the reversal operator we see that 
\begin{equation}
g(L_1^{m_1}L_2^{m_2}\dots L_q^{m_q}) = g(L_1^{m_q}L_2^{m_{q-1}}\dots L_q^{m_1}) = -(-1)^{\sum_i m_i} g(L_q^{m_1}L_{q-1}^{m_2}\dots L_1^{m_q}).
\end{equation}
Thus whenever $q$ is odd, (so that  $L_{q-i}=L_i$), and $|w|=n=\sum_i m_i$ is even, then the Goldberg coefficient must vanish~\cite{Goldberg}. For example $g(XY^2X)=g(X^2YX)= g(XYX^2) = 0$. More generally, for arbitrary non-negative integers $a$, $b$, $c$, we have
\begin{equation}
g(X^{2a+1}Y^{2b+2}X^{2c+1})=g(X^{2a+2}Y^{2b+1}X^{2c+1})= g(X^{2a+1}Y^{2b+1}X^{2c+2}) = 0.
\end{equation}
Identical results hold under the interchange $X \leftrightarrow Y$. Fundamentally, this symmetry is why so many of the Goldberg coefficients for \emph{even} length words are zero. 

Conversely, if we demand $g(w)\neq0$  for a word of even length $|w|=2n$ then a \emph{necessary} condition, (not a \emph{sufficient} condition), is that it must be possible to write $w =  L_1^{m_1}L_2^{m_2}\dots L_{2q}^{m_{2q}}$ with $L_i\neq L_{i+1}$ and $m_i\geq 1$ and $|w|=2n=\sum_i m_i$.

\item One of the few explicitly known evaluations of the Goldberg coefficients is~\cite{Goldberg}
\begin{equation}
g(X^aY^b) =  g(X^b Y^a) = (-1)^{a+b+1} g(Y^a X^b)= (-1)^{a+b+1} g(Y^b X^a) = {(-1)^{a}\over a!b!} \sum_{i=1}^b {b \choose i} \; {\bf B}_{a+b-i},
\end{equation}
where ${\bf B}_{a+b-i}$ are the usual Bernoulli numbers, in the usual convention where ${\bf B}_1 = -1/2$. In particular
\begin{equation}
g(X^aY) =  g(X Y^a) = (-1)^{a} g(Y^a X)= (-1)^{a} g(Y X^a) = {(-1)^{a}\over a!} \; {\bf B}_{a},
\end{equation}
Since in this convention all other odd Bernoulli numbers are zero, for $m\geq 1$ we have 
 \begin{equation}
g(X^{2m+1}Y) =  g(X Y^{2m+1}) = g(Y^{2m+1} X)=  g(Y X^{2m+1}) = 0,
\end{equation}
while
\begin{equation}
g(X^{2m}Y) =  g(X Y^{2m}) = g(Y^{2m} X)=  g(Y X^{2m}) =  {{\bf B}_{2m}\over(2m)!} \neq 0.
\end{equation}
\end{itemize}

\section{The number of non-zero Goldberg coefficients}
\label{S:bounds}

Let us denote by $\#_n$ the number of non-zero Goldberg coefficients on words of length $n$. That is
\begin{equation}
\#_n = \#\left\{ w:  g(w) \neq 0 \quad\hbox{and}\quad |w|=n \right\}.
\end{equation}
What can we say about these numbers?

\subsection{Bounds on $\#_n$}

Since there are exactly $2^n$ words of length $n$ on an alphabet of two letters, we certainly have $\#_n \leq 2^n$. But we can actually do better than that. For $n\geq 2$ the words $X^n$ and $Y^n$ never contribute to the Goldberg series, $g({X^n}) = 0 = g({Y^n})$, consequently for $n\geq 2$ we have $\#_n \leq 2^n -2$. Inspection of Table I quickly leads to the \emph{observation} that this inequality is actually saturated whenever $n$ is a prime number. 

\begin{table}[!h]
\caption{Number of non-zero Goldberg coefficients}
\begin{tabular}{|c|c|c|c|c|}
\hline
n & $\#_n$ & $2^n-2$ &$\#_n/(2^n-2)$& $\#_n/(2^n-2)$\\
\hline
2 & 2 & 2 & 1 & 1.0000\\
3 & 6 & 6& 1& 1.0000\\
4 & 4 & 14& 2/7& 0.2857\\
5& 30 & 30& 1& 1.0000\\
6 & 28 & 62& 14/31& 0.4516\\
7 & 126 & 126&1 &1.0000\\
8 & 124 & 254& 62/127& 0.4882\\
9 & 390 & 510 & 13/17& 0.7647\\
10& 388 &1022 & 194/511& 0.3796\\
11 & 2046 & 2046 & 1 & 1.0000\\
12 & 2044 & 4094 & 1022/2047 &0.4993\\
13& 8190 &8190 & 1 & 1.0000\\
14& 8188 & 16382 & 4094/8191 &0.4998\\
15 &29776 & 32766 & 4961/5461& 0.9804\\
16 & 30124 & 65534 & 15062/32767& 0.4597\\
17&131070 & 131070 & 1 &1.0000\\
\hline
\end{tabular}
\end{table}

To \emph{formally prove} this we proceed as follows. We start with the standard result
\begin{equation}
Z(X,Y) = \ln\left( e^X e^Y\right)  
 = \sum_{k>0} {(-1)^{k-1}\over k} \sum_{r_i+s_i > 0;  \atop1\leq i \leq k} 
  {X^{r_1} Y^{s_1} X^{r_2} Y^{s_2}  \cdots X^{r_k} Y^{s_k}\over r_1! s_1!  r_2! s_2! \cdots  r_k! s_k!}.
\end{equation}
Then
\begin{equation}
z_n(X,Y)= \sum_{k>0} {(-1)^{k-1}\over k} \sum_{r_i+s_i > 0; \; \; \sum_i (r_i+s_i) = n \atop1\leq i \leq k} 
  {X^{r_1} Y^{s_1} X^{r_2} Y^{s_2}  \cdots X^{r_k} Y^{s_k}\over r_1! s_1!  r_2! s_2! \cdots  r_k! s_k!}.
\end{equation}
Unfortunately the ``basis'' used above is over-complete. For example $X Y^2 X = X Y X^0 Y X$, etc. That is,  many different $X^{r_1} Y^{s_1} X^{r_2} Y^{s_2}  \cdots X^{r_k} Y^{s_k}$ can contribute to the same word $w(X,Y)$. 
To deal with this let us define
\begin{equation}
W\left(X^{r_1} Y^{s_1} X^{r_2} Y^{s_2}  \cdots X^{r_k} Y^{s_k}\right)  
= X^{\rho_1} Y^{\sigma_1} X^{\rho_2} Y^{\sigma_2}  \dots X^{\rho_K} Y^{\sigma_K}.
\end{equation}
Here we ``collapse'' the symbol string $X^{r_1} Y^{s_1} X^{r_2} Y^{s_2}  \dots X^{r_k} Y^{s_k}$ to generate a ``word''. We do this by eliminating any interior zeros, and then merging adjacent indices that correspond to the same letter. At worst the only remaining zero indices are $\rho_1$ and/or $\sigma_K$. That is,  by construction $\prod_{i=2}^K \rho_i \neq 0 \neq \prod_{i=1}^{K-1} \sigma_i$ and $K \leq k \leq n$.

The Goldberg coefficients are then
\begin{equation}
g(w) =  \sum_{k>0} {(-1)^{k-1}\over k} \sum_{r_i+s_i > 0; \; \; \sum_i (r_i+s_i) = |w| \atop1\leq i \leq k} 
 {\delta( w = W(X^{r_1} Y^{s_1} X^{r_2} Y^{s_2}  \dots X^{r_k} Y^{s_k}))\over r_1! s_1!  r_2! s_2! \dots  r_k! s_k!},
\end{equation}
with
\begin{equation}
|w| = \sum_{j=1}^K (\rho_j+\sigma_i) = \sum_{i=1}^k (r_i+s_i).
\end{equation}
Here $\delta(w_1 = w_2)$ is the ``indicator function'' which equals $1$ when $w_1=w_2$ and equals $0$ when $w_1\neq w_2$.

In the sum over $k$ we never get higher than $k=|w|$, nor lower than $k=K$, so in fact
\begin{equation}
g(w) =  \sum_{k=K}^{|w|}  {(-1)^{k-1}\over k} \sum_{r_i+s_i > 0; \;\; \sum_i (r_i+s_i) = |w|  \atop1\leq i \leq k}  
{\delta( w = W(X^{r_1} Y^{s_1} X^{r_2} Y^{s_2}  \dots X^{r_k} Y^{s_k}))
\over r_1! s_1!  r_2! s_2! \dots  r_k! s_k!}.
\label{E:explicit}
\end{equation}
(Proof: Obvious. Observe that  $|w| = \sum_i (r_i + s_i) \geq  \sum_i 1 = k \geq K$.)

Let us now assume that $|w|=p$ is a prime. In the sum for $g(w)$, separate out the contribution with the highest value of $k$, (explicitly this is $k=|w|=p$).
Then
\begin{eqnarray}
g({w: |w|=p}) &=& 
 {(-1)^{p-1}\over p}  \sum_{r_i+s_i > 0; \; \; \sum_i (r_i+s_i) = p \atop1\leq i \leq p} 
 {\delta( w = W(X^{r_1} Y^{s_1} X^{r_2} Y^{s_2}  \dots X^{r_p} Y^{s_p}))\over r_1! s_1!  r_2! s_2! \dots  r_p! s_p!} 
 \nonumber\\
 &+&
\sum_{k=K}^{p-1} {(-1)^{k-1}\over k} \sum_{r_i+s_i > 0; \; \; \sum_i (r_i+s_i) = p \atop1\leq i \leq k} 
 {\delta( w = W(X^{r_1} Y^{s_1} X^{r_2} Y^{s_2}  \dots X^{r_k} Y^{s_k}))\over r_1! s_1!  r_2! s_2! \dots  r_k! s_k!}.
\end{eqnarray}
But in that first term, $\forall i\in\{1,2,\dots,p\}$ we have $r_i+s_i=1$ so that $\{r_i,s_i\} = \{1,0\} \hbox{ or } \{0,1\}$; so the factorials trivialize. For fixed $w$ the remaining sum in that first term has only one contribution. 
That is
\begin{equation}
g({w: |w|=p}) =  {(-1)^{p-1}\over p}  + \sum_{k=K}^{p-1} {(-1)^{k-1}\over k} \sum_{r_i+s_i > 0; \; \; \sum_i (r_i+s_i) = p \atop1\leq i \leq k} 
 {\delta( w = W(X^{r_1} Y^{s_1} X^{r_2} Y^{s_2}  \dots X^{r_k} Y^{s_k}))\over r_1! s_1!  r_2! s_2! \dots  r_k! s_k!}.
\end{equation}
Now suppose that $X^p\neq w \neq Y^p$, then in the trailing terms we have $1< k < p$, while $r_i<p$ and $s_i<p$. So none of the trailing terms can individually have a factor $p$ in the denominator. Thus, when the trailing terms are all summed and reduced to lowest rational form, there cannot be a factor $p$ in the denominator. So there is nothing available to cancel the $1/p$ in the leading term. That is, for any word of prime length $g({w: |w|=p})\neq 0$, modulo the exceptional cases $X^p\neq w \neq Y^p$.
In those two exceptional cases $w=X^p$ or $w=Y^p$ we already know $g(w)=0$.
That is, whenever $|w|=p$ is a prime number, we have $\#_p = 2^p-2$; the number of non-zero Goldberg coefficients is maximal.

Consequently $\limsup (\#_n / 2^n) = 1$ and the number of terms in the Goldberg expansion grows exponentially with the word length. 
It is this observation that makes it absolutely clear that obtaining truly extensive tables of Goldberg coefficients is an intrinsically hopeless task.

\subsection{Bounds on $\#_{2n}$}
Furthermore, we note that whenever $n=p+1$ is one more than an odd prime, 
(that is, excluding the unique even prime 2), then \emph{observationally} it seems that we have $\#_{p+1} = 2^p - 4$. 
Let us first establish the bound that $\#_{2n} \leq 2^{2n-1}-4$ for $n\geq 2$.

Consider an arbitrary word $w_{2n-1}$ of length $2n-1$, and write it in the form $w_{2n-1} = L_1^{m_1}L_2^{m_2}\dots L_q^{m_q}$. Then two words of length $2n$ are obtained by appending either a $X$ or a $Y$:
\begin{equation}
w_{2n:X} = w_{2n-1} X = L_1^{m_1}L_2^{m_2}\dots L_q^{m_q} X; \qquad w_{2n:Y} = w_{2n-1} X = L_1^{m_1}L_2^{m_2}\dots L_q^{m_q} Y.
\end{equation}
Now either $g(w_{2n:X})=0$ or $g(w_{2n:Y})=0$. (If $L_1=X$, then $g(w_{2n:X})=0$;  if $L_1=Y$, then $g(w_{2n:Y})=0$.) Since there are $2^{2n-1}$ arbitrary words $w_{2n-1}$, then at least $2^{2n-1}$ of the $g(w_{2n})$ are zero. Additionally, for $n\geq 2$ we also know the four special cases $g(X^{2n-1}Y)=g(Y^{2n-1}X)=g(XY^{2n-1}) =g(YX^{2n-1})=0$, so at least $2^{2n-1}+4$ of the $g(w_{2n})$ are zero. Since there are $2^{2n}$ words of length $2n$, we see that at most $2^{2n-1}-4$ of them are non-zero: That is, $\#_{2n} \leq 2^{2n-1}-4$ for $n\geq 2$. Observationally it seems that this bound is saturated when $2n=p+1$ for $p$ an odd prime (that is, exclude $p=2$).

To \emph{formally prove} that  this bound actually saturates for $2n=p+1$ consider words of the form $w =  L_1^{m_1}L_2^{m_2}\dots L_{2q}^{m_{2q}}$, with $|w|=2n=\sum_i m_i$, which we have already seen are the only possibilities for $|w|$ even,  and write the relevant Goldberg coefficients as
\begin{equation}
g( L_1^{m_1}L_2^{m_2}\dots L_{2q}^{m_{2q}}) =
  \sum_{k\geq q}^{2n} {(-1)^{k-1}\over k} \sum_{r_i+s_i > 0; \; \; \sum_i (r_i+s_i) =2n \atop1\leq i \leq k} 
 {\delta( L_1^{m_1}L_2^{m_2}\dots L_{2q}^{m_{2q}} = W(X^{r_1} Y^{s_1} X^{r_2} Y^{s_2}  \dots X^{r_k} Y^{s_k}))
 \over r_1! s_1!  r_2! s_2! \dots  r_k! s_k!}.
\end{equation}
Specialize to $2n=p+1$, and separate out the top two terms
\begin{eqnarray}
g( L_1^{m_1}L_2^{m_2}\dots L_{2q}^{m_{2q}}) &=&
  {(-1)^{p}\over p+1} \sum_{r_i+s_i > 0; \; \; \sum_i (r_i+s_i) =p+1 \atop1\leq i \leq p+1} 
 {\delta( L_1^{m_1}L_2^{m_2}\dots L_{2q}^{m_{2q}} = W(X^{r_1} Y^{s_1} X^{r_2} Y^{s_2}  \dots X^{r_{p+1}} Y^{s_{p+1}}))
 \over r_1! s_1!  r_2! s_2! \dots  r_{p+1}! s_{p+1}!}
 \nonumber\\
 &+&
 {(-1)^{p-1}\over p} \sum_{r_i+s_i > 0; \; \; \sum_i (r_i+s_i) =p+1 \atop1\leq i \leq p} 
 {\delta( L_1^{m_1}L_2^{m_2}\dots L_{2q}^{m_{2q}} = W(X^{r_1} Y^{s_1} X^{r_2} Y^{s_2}  \dots X^{r_p} Y^{s_p}))
 \over r_1! s_1!  r_2! s_2! \dots  r_p! s_p!}
 \nonumber\\
&+&
 \sum_{k\geq q}^{2p-1} {(-1)^{k-1}\over k} \sum_{r_i+s_i > 0; \; \; \sum_i (r_i+s_i) =p+1 \atop1\leq i \leq k} 
 {\delta( L_1^{m_1}L_2^{m_2}\dots L_{2q}^{m_{2q}} = W(X^{r_1} Y^{s_1} X^{r_2} Y^{s_2}  \dots X^{r_k} Y^{s_k}))
 \over r_1! s_1!  r_2! s_2! \dots  r_k! s_k!}.
 \nonumber\\
\end{eqnarray}
In the first line the constraints imply $\forall i$ that $r_i+s_i=1$, whence either $r_i=1$ and $s_i=0$, or $r_i=0$ and $s_i=1$. All the factorials trivialize and the remaining sum reduces to unity. Since $p$ is an odd prime ($p\neq 2$) we now have
\begin{eqnarray}
g( L_1^{m_1}L_2^{m_2}\dots L_{2q}^{m_{2q}}) &=&
  -{1\over p+1} 
 \nonumber\\
 &+&
 {1\over p} \;\;\;\sum_{r_i+s_i > 0; \; \; \sum_i (r_i+s_i) =p+1 \atop1\leq i \leq p} 
 {\delta( L_1^{m_1}L_2^{m_2}\dots L_{2q}^{m_{2q}} = W(X^{r_1} Y^{s_1} X^{r_2} Y^{s_2}  \dots X^{r_p} Y^{s_p}))
 \over r_1! s_1!  r_2! s_2! \dots  r_p! s_p!}
 \nonumber\\
&+&
\sum_{k\geq q}^{2p-1} {(-1)^{k-1}\over k} \sum_{r_i+s_i > 0; \; \; \sum_i (r_i+s_i) =p+1 \atop1\leq i \leq k} 
 {\delta( L_1^{m_1}L_2^{m_2}\dots L_{2q}^{m_{2q}} = W(X^{r_1} Y^{s_1} X^{r_2} Y^{s_2}  \dots X^{r_k} Y^{s_k}))
 \over r_1! s_1!  r_2! s_2! \dots  r_k! s_k!}.
 \nonumber\\
\end{eqnarray}
In the second line, the constraints imply that for all but one of the $i$, (say for $i\neq i_*$),  we have $r_i+s_i=1$, and that for exactly one of the $i$ this quantity equals two, say $r_{i_*}+s_{i_*}=2$. Then the coefficient multiplying the $1/p$ in the second line above is
\begin{equation}
\sum_{i_*=1}^p \;\;\;\sum_{r_i+s_i =1; \;r_{i_*}+r_{s_*}=2 \;\; \atop1\leq i \leq p} 
 {\delta( L_1^{m_1}L_2^{m_2}\dots L_{2q}^{m_{2q}} = W(X^{r_1} Y^{s_1} X^{r_2} Y^{s_2}  \dots X^{r_p} Y^{s_p}))
 \over r_{i_*}! s_{i_*}!}.
\end{equation}
This collapses to
\begin{equation}
\sum_{i_*=1}^p \left\{ 
{1\over2}  \delta\left( L_{i_*} L_{i_*+1} = X^2\right)
+
\delta\left( L_{i_*} L_{i_*+1} = XY  \right)
+
{1\over2}  \delta\left( L_{i_*} L_{i_*+1} =  Y^2 \right) 
\right\}.
\end{equation}
Here, (with a minor change of notation), $ L_{i_*} L_{i_*+1}$ are the two letters at positions ${i_*} $ and ${i_*+1}$ when the word $L_1^{m_1}L_2^{m_2}\dots L_{2q}^{m_{2q}} \to L_1 L_2 \dots L_{2n}$ is expanded out in full. 
This sum now collapses to
\begin{equation}
{1\over2} \left[ (2n-1) -(2q-1)\right] + q = {1\over2} p - q + {1\over2} + q = {p+1\over2}.
\end{equation}
(Note that the final result for this sum is independent of whether the word begins with $X$ and ends with $Y$, or begins with $Y$ and ends with $X$.)
So at this stage we have
\begin{eqnarray}
g( L_1^{m_1}L_2^{m_2}\dots L_{2q}^{m_{2q}}) &=&
  -{1\over p+1} + {1\over2} + {1\over2p}
 \nonumber\\
&+&
\sum_{k\geq q}^{2p-1} {(-1)^{k-1}\over k} \sum_{r_i+s_i > 0; \; \; \sum_i (r_i+s_i) =p+1 \atop1\leq i \leq k} 
 {\delta( L_1^{m_1}L_2^{m_2}\dots L_{2q}^{m_{2q}} = W(X^{r_1} Y^{s_1} X^{r_2} Y^{s_2}  \dots X^{r_k} Y^{s_k}))
 \over r_1! s_1!  r_2! s_2! \dots  r_k! s_k!}.
 \nonumber\\
\end{eqnarray}

Consider the cases:
\begin{itemize}
\item 
If $q\geq 3$, then certainly $k\geq 3$, and then each of the $r_i<p$ and $s_i<p$. Then this last line is a sum of rational numbers with no factors of $p$ in the denominator; when summed and reduced to lowest form this will be some rational number with no factors of $p$ in the denominator. 
\item
 If $q=2$ then $k\geq 2$. The terms with $k\geq 3$ again have $r_i<p$ and $s_i<p$. Among the terms with $k=2$, only those with $r_i<p$ and $s_i<p$ contribute to the sum. Then this last line is again a sum of rational numbers with no factors of $p$ in the denominator; when summed and reduced to lowest form this will be some rational number with no factors of $p$ in the denominator. 
\end{itemize} 
 So for $q\geq 2$, (remembering $\sum_i m_i = 2n = p+1$),  we have
\begin{equation}
g( L_1^{m_1}L_2^{m_2}\dots L_{2q}^{m_{2q}}) = {1\over2p}  + (\hbox{rational number with no $p$'s in denominator}) \neq 0.
\end{equation}
For $q=1$ the only possible contributions that have $r_i=p$ or $s_i = p$,  (and so have any chance at all of summing to zero),   are those special cases that we knew were zero anyway, which were explicitly excluded by putting the ``$-4$'' in the bound. 
In brief, $\#_{2n}\leq 2^{2n-1}-4$, and this bound is saturated whenever $2n=p+1$ with $p$ an odd prime ($p\neq 2$): $\#_{p+1} = 2^p - 4$. 
Consequently $\limsup(\#_{2n}/2^{2n}) = 1/2$.

\subsection{Comment}

This is as much as we have been able to do in terms of placing explicit bounds on $\#_n$ and in terms of understanding the pattern of zeros of the Goldberg coefficients $g(w)$. We fully expect that more could in principle be said. 
\\
(For instance: It would be nice to know if  $\#_{n}/(2^n-2)$ is locally minimum whenever $n=p+1$? Or whether or not $\liminf (\#_n/2^n) = 1/2$?)

\section{Symmetric Baker--Campbell--Hausdorff formula} 

Let us now consider the ``symmetric'' version of the Baker--Campbell--Hausdorff formula
\begin{equation}
S(X,Y)=\ln\left( e^{X/2} e^Y e^{X/2} \right) = \sum_{n=1}^\infty s_n(X,Y).
\end{equation}
Following along the lines of the brief discussion by Reinsch~\cite{Reinsch}, 
our simplified Reinsch algorithm can just as easily be adapted to investigating this series.
First note that, since $\left( e^{X/2} e^Y e^{X/2} \right)^{-1}=  e^{-X/2} e^{-Y} e^{-X/2}$ we have $S(X,Y)=-S(-X,-Y)$, which implies the well-known result that all the even-level terms vanish: $s_{2n}(X,Y)=0$. Killing off half the terms in the expansion is definitely a worthwhile simplification.  

To reconstruct the standard Baker--Campbell--Hausdorff formula one can, (using the usual notation $L_X Y= [X,Y] = \mathrm{ad}_X Y$; the $L_X Y$ (``Lie bracket'') notation is more common on the physics community, the $ \mathrm{ad}_X Y$ (``adjoint'') notation is more common within the mathematics community), adapt the  Baker--Hausdorff lemma
\begin{equation}
e^X Y e^{-X} = \exp(L_X) Y = \sum_{n=0}^\infty {L_X^n\over n!}\; Y  = \sum_{n=0}^\infty {[X^nY]\over n!} = \left[ e^X Y \right] ,
\end{equation}
to see that
\begin{equation}
e^X e^Y 
=  e^{X/2} \left(e^{X/2} e^Y e^{-X/2}\right) e^{X/2} 
= e^{X/2} \left(e^{(e^{X/2} Y e^{-X/2})} \right) e^{X/2}
=e^{X/2} \left(e^{\exp(L_{X/2})Y} \right) e^{X/2}.
\end{equation}
That is, the usual Baker--Campbell--Hausdorff and symmetric Baker--Campbell--Hausdorff formulae are related by
\begin{equation}
Z(X,Y) = S(X, {\exp(L_{X/2})Y}).
\end{equation}
Similarly
\begin{equation}
Z(X,Y) = \exp(L_{X/2}) S(X, Y); \qquad\hbox{and}\qquad S(X,Y) = \exp(-L_{X/2}) Z(X, Y).
\end{equation}

Let us now compute (eg, using {\sf Maple} or some equivalent package), the $(N+1)\times(N+1)$ matrix
\begin{equation}
S_N = \ln\left( \exp(X_N/2) \exp(Y_N) \exp(X_N/2)  \right).
\end{equation}
Again, the first row of the matrix $S_n$ is (essentially) the information we want. Specifically, noting that the matrix $S_N$ is strictly upper triangular, let us denote
\begin{equation}
S_N = 
\left[ \begin{array}{ccccccc}  
0&s_1&s_2&s_3&\cdots &s_{N-1} &s_N \\
0&0&*&*&\cdots &* &*     \\
0&0&0&*&\cdots &* &*     \\
\vdots &\vdots&\vdots& \vdots & \ddots &\vdots& \vdots\\
0&0&0&0&\cdots & *&*    \\
0&0&0&0&\cdots  &0& *     \\
0&0&0&0&\cdots & 0  &0
\end{array}\right].
\end{equation}
Here the $*$'s again denote nonzero quantities that are not of specific interest for current purposes.  It is now a trivial exercise, (see the {\sf Maple} code in the Appendix), to obtain the first few terms
\begin{eqnarray}
s_1 &=& x_1 + y_1;
\\
s_2 &=& 0;
\\
s_3 &=& {1\over24}(2x_1 y_2 y_3 - x_1 x_2 y_3 -4y_1 x_2 y_3 -y_1x_2 x_3+2 y_1 y_2 x_3 +2x_1 y_2 x_3);
\\
s_4 &=&0.
\end{eqnarray}
Again, higher-order  terms are easy to calculate, but tedious to display.
Converting these low-order terms to words the key observations are that 
\begin{equation}
s_1(X,Y) = X+Y; \qquad
s_3(X,Y) = {1\over24}(2XY^2 - X^2Y -4YXY -YX^2+2 Y^2X +2XYX).
\end{equation}
Converting to commutators, (which for higher orders would be a very labour intensive process), we have
\begin{equation}
s_3(X,Y) = -{1\over24}[(X+2Y)XY].
\end{equation}
We could define ``symmetric'' versions of the Goldberg coefficients, $g_{\scriptscriptstyle{S}}(w)$, and proceed with a fuller analysis along the lines above, but have not yet done so.
We content ourselves (see Table II) with calculating $\#_n^S$ for the symmetric Baker--Campbell--Hausdorff expansion and comparing it with our explicit bound. Note $\#_{2n}^S = 0$, and that $\#_{2n+1}^S$ again saturates whenever $2n+1=p$ is an odd prime. (The only even prime, 2, has to be treated separately since now $\#_2^S=0$.)

\begin{table}[!h]
\caption{Number of non-zero Goldberg coefficients
 (symmetric BCH)}
\begin{tabular}{|c|c|c|c|c|}
\hline
$n$ & $\#_n^S$ & $2^n-2$ &$\#_n^S/(2^n-2)$& $\#_n^S/(2^n-2)$\\
\hline
3 & 6 & 6& 1& 1.0000\\
5& 30 & 30& 1& 1.0000\\
7 & 126 & 126&1 &1.0000\\
9 & 435 & 510 & 29/34& 0.8529\\
11 & 2046 & 2046 & 1 & 1.0000\\
13& 8190 &8190 & 1 & 1.0000\\
15 & 30846 & 32766 & 15423/16388& 0.9411\\
17&131070 & 131070 & 1 &1.0000\\
\hline
\end{tabular}
\end{table}

\section{Some other variants of the Baker--Campbell--Hausdorff formula} 

Now that we have seen the basic techniques employed, many other variants of  the Baker--Campbell--Hausdorff formula can easily be analyzed and appropriate series extracted by using our simplified Reinsch algorithm.
For instance, consider the variants presented below. (For this Section, we will not go beyond 4$^{th}$ order in explicit formulae.) These are all specific instances of the general fact that the simplified Reinsch algorithm can easily be applied to any function $f(X,Y)$ of non-commuting variables $X$ and $Y$; if it is known that a formal power series can be constructed, then the simplified Reinsch algorithm will find it. 
For example:
\begin{itemize}
\item Consider a multiple product 
\begin{equation}
P(X_i) = \ln\left(\prod_{i=1}^m \exp(X_i)\right) = \sum_{n=1}^\infty p_n(X_i),
\end{equation}
then by iterating the usual BCH expansion, it is easy to see that the two lowest-order terms in this generalized BCH expansion are
\begin{equation}
p_1(X_i) = \sum_{i=1}^n X_i;
\qquad\hbox{and}\qquad
p_2(X_i) = {1\over2} \sum_{i=1}^{n-1}\sum_{j=i+1}^n [X_i,X_j].
\end{equation}
\item 
Consider the palindromic multiple product 
\begin{equation}
Q(X_i) = \ln\left(\prod_{i=1}^m \exp(X_i) \; \prod_{i=m}^1 \exp(X_i)\right) = \sum_{n=1}^\infty q_n(X_i),
\end{equation}
where the two multiple products are oppositely ordered but otherwise identical. (For instance $S(X,Y)=\ln\left( e^{X/2} e^Y e^{X/2} \right)$ can be rewritten as $S(X,Y)  =\ln\left( e^{X/2} e^{Y/2} \; e^{Y/2} e^{X/2} \right)$ which is manifestly of this form.)  Then by considering inverses we have 
\begin{equation}
-Q(X_i) = \ln\left(\prod_{i=1}^m \exp(-X_i) \; \prod_{i=m}^1 \exp(-X_i)\right) = Q(-X_i),
\end{equation}
implying
\begin{equation}
- q_n(X_i) = q_n(-X_i) = (-1)^n q_n(X_i).
\end{equation}
Consequently all of the even-order terms are zero, $q_{2n}(X_i)=0$ for any of these palindromic versions of BCH. 
\end{itemize}
By applying one or both of these observations and suitably choosing the $X_i$ we can often arrange low-order terms to vanish, which sometimes leads to interesting modified BCH relations.

\subsection{Loop Baker--Campbell--Hausdorff formula} 
\enlargethispage{50pt}
 Consider the ``loop'' Baker--Campbell--Hausdorff formula
\begin{equation}
L(X,Y)=\ln\left( e^{X} e^Y e^{-X} e^{-Y}\right) = \sum_{n=1}^\infty \ell_n(X,Y).
\end{equation}
This particular type of series is of interest, for instance, when considering differential geometric parallel transport around a closed curve, when trying to establish non-Abelian analogues of Stokes' theorem, or when considering Wigner rotation/Thomas precession in special relativity.
Then
\begin{eqnarray}
\ell_1 &=& 0;
\\
\ell_2 &=& x_1 y_2 - y_1 x_2;
\\
\ell_3 &=& {1\over2} \left( x_{{1}}x_{{2}}y_{{3}}-2\,x_{{1}}y_{{2}}x_{{3}}-x_{{1}}y_{{2}}y_{{3}}
+y_{{1}}x_{{2}}x_{{3}}+2\,y_{{1}}x_{{2}}y_{{3}}-y_{{1}}y_{{2}}x_{{3}}
\right);
\\
\ell_4 &=&{1\over12}\left(
2\,x_{{1}}x_{{2}}x_{{3}}y_{{4}}-6\,x_{{1}}x_{{2}}y_{{3}}x_{{4}}-3\,x_{{1}}x_{{2}}y_{{3}}y_{{4}}
+6\,x_{{1}}y_{{2}}x_{{3}}x_{{4}}+6\,x_{{1}}y_{{2}}x_{{3}}y_{{4}}+2\,x_{{1}}y_{{2}}y_{{3}}y_{{4}}\right.
\nonumber\\
&&\quad 
\left. 
-2\,y_{{1}}x_{{2}}x_{{3}}x_{{4}}-6\,y_{{1}}x_{{2}}y_{{3}}x_{{4}}-6\,y_{{1}}x_{{2}}y_{{3}}y_{{4}}
+3\,y_{{1}}y_{{2}}x_{{3}}x_{{4}}+6\,y_{{1}}y_{{2}}x_{{3}}y_{{4}}-2\,y_{{1}}y_{{2}}y_{{3}}x_{{4}}
\right).
\end{eqnarray}
Notice the $\ell_1$ term has quietly cancelled. The first few non-zero  terms (in word form) become
\begin{eqnarray}
\ell_2(X,Y) &=& XY-YX;
\\
\ell_3(X,Y) &=& {1\over2} \left(X^2 Y - 2 XYX - XY^2 + YX^2 + 2 YXY-Y^2 X
\right);
\\
\ell_4(X,Y) &=&{1\over12}\left(
2 X^3 Y - 6 X^2YX - 3 X^2 Y^2 +6XYX^2 +6 XYXY +2 XY^3
\right.
\nonumber\\
&&\quad 
\left. 
-2YX^3-6YXYX-6YXY^2+3Y^2X^2+6Y^2XY-2Y^3X
\right).
\end{eqnarray}
When converting to commutators, (which for higher orders would be an even more labour intensive process), we make use of the results
\begin{eqnarray}
&&
[X^2 Y - 2 XYX - XY^2 + YX^2 + 2 YXY-Y^2 X] = [X^2 Y - 2 XYX  + 2 YXY-Y^2 X]
\nonumber\\
&&\qquad\qquad\qquad
=[(X+2X+2Y+Y)XY] = 3[(X+Y)XY],
\end{eqnarray}  
and
\begin{eqnarray}
&&[2 X^3 Y - 6 X^2YX - 3 X^2 Y^2 +6XYX^2 +6 XYXY +2 XY^3
\nonumber\\
&&
\quad -2YX^3-6YXYX-6YXY^2+3Y^2X^2+6Y^2XY-2Y^3X]
\nonumber\\
&&\qquad\qquad\qquad
= [2 X^3 Y - 6 X^2YX  +6 XYXY -6YXYX+6Y^2XY-2Y^3X]
\nonumber\\
&&\qquad\qquad\qquad
= [ (2X^2+6X^2+6XY+6YX+6Y^2+2Y^2) XY ]
\nonumber\\
&&\qquad\qquad\qquad
= [(8X^2+6XY+6YX+8Y^2)XY]
\nonumber\\
&&\qquad\qquad\qquad
= 4[(2X^2+3XY+2Y^2)XY],
\end{eqnarray}
to deduce
\begin{eqnarray}
\ell_2(X,Y) &=& [XY];
\\
\ell_3(X,Y) &=& {1\over2} [(X+Y)XY];
\\
\ell_4(X,Y) &=&{1\over12} [(2X^2+3XY+2Y^2)XY]. 
\end{eqnarray}
Higher-order terms are (at least in principle) straightforward, though often it is the lowest-order non-zero term that is most useful
\begin{equation}
L(X,Y)=\ln\left( e^{X} e^Y e^{-X} e^{-Y}\right)  = [X,Y] + \mathcal{O}(X^2Y,XY^2). 
\end{equation}

\subsection{Triangular Baker--Campbell--Hausdorff formula} 
Now consider the ``triangular'' Baker--Campbell--Hausdorff formula
\begin{equation}
T(X,Y)=\ln\left( e^{-X} e^{(X+Y)} e^{-Y}\right) = \sum_{n=1}^\infty t_n(X,Y).
\end{equation}
(This ``triangular'' form will inherit many of the properties of the ``loop'' form considered above, it is perhaps just a little simpler.)~Then
\begin{eqnarray}
t_1 &=& 0;
\\
t_2 &=& -{1\over2}\left(x_1 y_2 - y_1 x_2\right);
\\
t_3 &=& {1\over6} \left( 
x_{{1}}x_{{2}}y_{{3}}-2\,x_{{1}}y_{{2}}x_{{3}}+x_{{1}}y_{{2}}y_{{3}}+y_{{1}}x_{{2}}x_{{3}}-2\,y_{{1}}x_{{2}}y_{{3}}+y_{{1}}y_{{2}}x_{{3}}
\right);
\\
t_4 &=&{1\over24}\left(
-x_{{1}}x_{{2}}x_{{3}}y_{{4}}+3\,x_{{1}}x_{{2}}y_{{3}}x_{{4}}-x_{{1}}x_{{2}}y_{{3}}y_{{4}}
-3\,x_{{1}}y_{{2}}x_{{3}}x_{{4}}+2\,x_{{1}}y_{{2}}x_{{3}}y_{{4}}-x_{{1}}y_{{2}}y_{{3}}y_{{4}}
\right.
\nonumber\\
&&\quad \quad
\left. 
+y_{{1}}x_{{2}}x_{{3}}x_{{4}}-2\,y_{{1}}x_{{2}}y_{{3}}x_{{4}}+3\,y_{{1}}x_{{2}}y_{{3}}y_{{4}}
+y_{{1}}y_{{2}}x_{{3}}x_{{4}}-3\,y_{{1}}y_{{2}}x_{{3}}y_{{4}}+y_{{1}}y_{{2}}y_{{3}}x_{{4}}
\right).
\end{eqnarray}
Notice the $t_1$ term has again quietly cancelled. The first few non-zero terms (in word form) become
\begin{eqnarray}
t_2(X,Y) &=& -{1\over2}\left(XY-YX\right);
\\
t_3(X,Y) &=& {1\over6} \left( 
X^2Y-2XYX+XY^2+YX^2-2YXY+Y^2X
\right);
\\
t_4(X,Y) &=&{1\over24}\left(
- X^3 Y+3 X^2YX-X^2Y^2-3XYX^2+2XYXY-XY^3
\right.
\nonumber\\
&&\quad \quad
\left. 
+YX^3-2YXYX+3YXY^2+Y^2X^2-3Y^2XY+Y^3X
\right).
\end{eqnarray}
Converting to commutators we obtain
\begin{eqnarray}
t_2(X,Y) &=& -{1\over2}[XY];
\\
t_3(X,Y) &=& {1\over6} [(X-Y) XY];
\\
t_4(X,Y) &=&-{1\over24} [(X^2-XY+Y^2) XY].
\end{eqnarray}
Higher-order terms are in principle straightforward, though often it is the lowest-order non-zero term that is most useful
\begin{equation}
T(X,Y)=\ln\left( e^{-X} e^{(X+Y)} e^{-Y}\right)   = -{1\over2}[X,Y] + \mathcal{O}(X^2Y,XY^2). 
\end{equation}

\subsection{Sum and difference Baker--Campbell--Hausdorff formula} 
Consider the ``sum and difference'' Baker--Campbell--Hausdorff formula
\begin{equation}
SD(X,Y)=\ln\left( e^{(X+Y)}  e^{(X-Y)}\right) = \sum_{n=1}^\infty sd_n(X,Y).
\end{equation}
The first few terms are
\begin{eqnarray}
sd_1 &=& 2x_1;
\\
sd_2 &=& -(x_1y_2-y_1x_2);
\\
sd_3 &=& {1\over3} \left( x_1 y_2 y_3 - 2 y_1 x_2 x_3 + y_1 y_2 x_3 \right);
\\
sd_4 &=& {1\over12} \left( 
x_{{1}}x_{{2}}x_{{3}}y_{{4}}-3\,x_{{1}}x_{{2}}y_{{3}}x_{{4}}+3\,x_{{1}}y_{{2}}x_{{3}}x_{{4}}-x_{{1}}y_{{2}}y_{{3}}y_{{4}}
\right.
\nonumber\\
&&\qquad\left.
-y_{{1}}x_{{2}}x_{{3}}x_{{4}}+3\,y_{{1}}x_{{2}}y_{{3}}y_{{4}}-3\,y_{{1}}y_{{2}}x_{{3}}y_{{4}}+y_{{1}}y_{{2}}y_{{3}}x_{{4}}
 \right)  .
\end{eqnarray}
These terms (in word form) become
\begin{eqnarray}
sd_1(X,Y) &=& 2X;
\\
sd_2(X,Y) &=& -(XY-YX);
\\
sd_3(X,Y) &=& {1\over3} \left(XY^2-2YX^2+Y^2X \right);
\\
sd_4(X,Y) &=& {1\over12} \left( 
X^3Y -3X^2YX+3XYX^2-XY^3- Y X^3 + 3 YXY^2 -3 Y^2XY +Y^3 X
 \right)  .
\end{eqnarray}
In commutator form this yields
\begin{eqnarray}
sd_1(X,Y) &=& 2X;
\\
sd_2(X,Y) &=& -[XY];
\\
sd_3(X,Y) &=& {1\over3} [Y^2X];
\\
sd_4(X,Y) &=& {1\over12} [(X^2-Y^2)XY].
\end{eqnarray}
Higher-order terms are in principle straightforward. (We had originally hoped, based on the fact that the third order term contains only one nested commutator, that this sum-and-difference form would be particularly simple, 
it is our melancholy duty to report that it is not.)

\subsection{Highly-symmetrized Baker--Campbell--Hausdorff formula} 
 Consider the ``highly-symmetrized'' Baker--Campbell--Hausdorff formula
\begin{equation}
SS(X,Y)=\ln\left( e^{-(X+Y)/2} e^{X/2} e^{Y} e^{X/2} e^{-(X+Y)/2}\right) = \sum_{n=1}^\infty ss_n(X,Y).
\end{equation}
This particular object has been carefully constructed to make \emph{as many as possible} of the low-order terms vanish. Specifically \emph{all} the $ss_{2n}=0$, \emph{and}  the $ss_1$ term has again quietly cancelled. The first few terms are
\begin{eqnarray}
ss_1 &=& 0;
\\
ss_2 &=& 0;
\\
ss_3 &=& {1\over24} \left( 
-x_{{1}}x_{{2}}y_{{3}}+2\,x_{{1}}y_{{2}}x_{{3}}+2\,x_{{1}}y_{{2}}y_{{3}}-y_{{1}}x_{{2}}x_{{3}}-4\,y_{{1}}x_{{2}}y_{{3}}+2\,y_{{1}}y_{{2}}x_{{3}}
\right);
\\
ss_4 &=& 0.
\end{eqnarray}
Among the first four terms (in word form) the only no-zero term is
\begin{eqnarray}
ss_3(X,Y) &=& {1\over24} \left( 
-X^2Y +2XYX+2XY^2-YX^2-4YXY+2Y^2X
\right).
\end{eqnarray}
In commutator form this becomes
\begin{eqnarray}
ss_3(X,Y) &=& -{1\over24} [(X+2Y)XY].
\end{eqnarray}
Higher-order terms are in principle straightforward. 

\subsection{Symmetric sum and difference Baker--Campbell--Hausdorff formula} 
Consider the ``symmetric sum and difference'' Baker--Campbell--Hausdorff formula
\begin{equation}
SSD(X,Y)=\ln\left( e^{(X-Y)/2} e^{(X+Y)}  e^{(X-Y)/2} \right) = \sum_{n=1}^\infty ssd_n(X,Y).
\end{equation}
This particular object has been carefully constructed to make \emph{as many as possible} of the low-order terms vanish.
The first few terms are
\begin{eqnarray}
ssd_1 &=& 2 x_1;
\\
ssd_2 &=& 0;
\\
ssd_3 &=& {1\over12} \left( 
-3\,x_{{1}}x_{{2}}y_{{3}}+6\,x_{{1}}y_{{2}}x_{{3}}+x_{{1}}y_{{2}}y_{{3
}}-3\,y_{{1}}x_{{2}}x_{{3}}-2\,y_{{1}}x_{{2}}y_{{3}}+y_{{1}}y_{{2}}x_{
{3}}
\right);
\\
ssd_4 &=& 0.
\end{eqnarray}
Among the first four terms (in word form) the only no-zero ones are
\begin{eqnarray}
ssd_1(X,Y) = 2X; 
\qquad\qquad
ssd_3(X,Y) &=& {1\over12} \left( 
-3 X^2 Y+6XYX +XY^2 - 3 YX^2 -2 YXY + Y^2 X
\right).
\end{eqnarray}
Then in commutator form
\begin{eqnarray}
ssd_3(X,Y) &=& -{1\over12}[(3X+Y)XY].
\end{eqnarray}
Higher-order terms are in principle straightforward. 

\subsection{Highly-symmetrized sum and difference Baker--Campbell--Hausdorff formula} 
Consider the ``highly-symmetrized sum and difference'' Baker--Campbell--Hausdorff formula
\begin{equation}
SSSD(X,Y)=\ln\left( e^{-X} e^{(X-Y)/2} e^{(X+Y)}  e^{(X-Y)/2} e^{-X}\right) = \sum_{n=1}^\infty sssd_n(X,Y).
\end{equation}
Again, most of the low-order terms vanish. 
The first few terms are
\begin{eqnarray}
sssd_1 &=& 0;
\\
sssd_2 &=& 0;
\\
sssd_3 &=& {1\over12} \left( 
-3\,x_{{1}}x_{{2}}y_{{3}}+6\,x_{{1}}y_{{2}}x_{{3}}+x_{{1}}y_{{2}}y_{{3}}-3\,y_{{1}}x_{{2}}x_{{3}}-2\,y_{{1}}x_{{2}}y_{{3}}+y_{{1}}y_{{2}}x_{{3}}
\right);
\\
sssd_4 &=& 0.
\end{eqnarray}
(Note that only the first term changes compared to  the previous ``symmetric sum and difference'' result.)
Among the first four terms (in word form) the only no-zero one is
\begin{eqnarray}
sssd_3(X,Y) &=& {1\over12} \left( 
-3 X^2 Y+6XYX +XY^2 - 3 YX^2 -2 YXY + Y^2 X
\right).
\end{eqnarray}
Then in commutator form
\begin{eqnarray}
sssd_3(X,Y) &=& -{1\over12}[(3X+Y)XY].
\end{eqnarray}
Higher-order terms are in principle straightforward. 

\subsection{Comment}

One general message to take from all these variants on the Baker--Campbell--Hausdorff expansion 
is that it is often possible to make several of the terms in the expansion vanish --- at the cost of making the quantity being expanded more complicated.  
The second general message is that, once one has developed and understood the simplified Reinsch algorithm for the ``standard'' Baker--Campbell--Hausdorff 
expansion, the simplified Reinsch algorithm can very easily be recycled to analyze many related matrix logarithms.
Indeed the simplified Reinsch algorithm can easily be applied to any function $f(X,Y)$ of non-commuting variables $X$ and $Y$; if it is somehow known that a formal power series can be constructed, then the simplified Reinsch algorithm will find it.

\section{Discussion}

We have provided a simplified version of the Reinsch algorithm for the Baker--Campbell--Hausdorff expansion and, (using completely standard ``off the shelf'' symbolic manipulation software), used it to investigate the properties of the Goldberg coefficients.  
Some suggestive patterns and bounds, \emph{observationally} motivated by inspecting the low-order and medium-high-order terms in the expansion, were then verified by analytic proof. 
We have also considered various variant forms of the Baker--Campbell--Hausdorff expansion, and using appropriate modifications of the simplified Reinsch algorithm, inspected the low-order terms. We would argue that one of the the main virtues of the 
simplified version of the Reinsch algorithm lies in the fact that it makes it almost trivial to carry out low-order ``symbolic experiments'', 
which can then be used to observationally suggest questions that might be amenable to direct analytic investigation.

\acknowledgments
This research was supported by the Marsden Fund, through a grant administered by the Royal Society of New~Zealand. AVB was also supported by a 2014--2015 Victoria University of Wellington Summer Scholarship.
We also wish to thank the referee for identifying a number of errors, and some obscurities in the presentation.

\clearpage
\appendix
\section*{Appendix: Some {\sf Maple} code}
\label{A:Maple}

This fragment of {\sf Maple} code will calculate the $z_n$. (For definiteness, the first 9 of the $z_n$).

\smallskip
\noindent
{\tt  
restart:\\
with(LinearAlgebra):\\
N:=9:\\
X:=Matrix(N+1,N+1):\\
Y:=Matrix(N+1,N+1):\\
for i from 1 to N do: X[i,i+1]:=x[i]: od:\\
for i from 1 to N do: Y[i,i+1]:=y[i]: od:\\
EX:=MatrixExponential(X):\\
EY:=MatrixExponential(Y):\\
EXY:=EX.\!EY:\\
Z:=MatrixFunction(EXY, ln(x), x):\\
for n from 1 to N do: z[n]:= Z[1,n+1]; od:
}

\smallskip
\noindent
The most time consuming part of the process is writing the results to the screen.

\smallskip
\noindent
{\tt  
for n from 1 to N do: z[n]; od;
}

\smallskip
\noindent
The first four terms ($z_1$, $z_2$, $z_3$, $z_4$), and some of the medium-low-order terms ($z_5$, $z_6$,  $z_7$, and $z_8$) are presented in full above. 
Modern laptops can easily calculate out to $n=13$, and with a little more difficulty out to $n=17$. The relevance of this {\sf Maple} code is its extreme simplicity, using completely ``off the shelf'' software --- we specifically do not make any claims regarding efficiency. 

\smallskip
\noindent
To count the number of non-zero Goldberg coefficients on words of length $n$, and compare it with the maximum possible value of $2^n-2$,  one can proceed as follows:

\smallskip
\noindent
{\tt  
for n from 2 to N do:
\quad 
print("n =", n, "\verb?#_n =?", nops(z[n]), "\verb#2^n-2# =", \verb#2^n-2# ); 
\quad
od;
}

\smallskip
\noindent
By \emph{not} explicitly printing the $z_n$ to screen, one saves a lot of time, and the calculation can be carried out to somewhat higher order. 

\smallskip
\noindent
This code, or minor variants thereof, is sufficient for one to be able to easily check the first few terms in this or closely related computations.
For instance, the changes required to calculate the $s_n$ appearing in symmetric object $S(X,Y) = \ln\left(e^{X/2} e^Y e^{X/2}\right)$ are straightforward.

\smallskip
\noindent
{\tt  
EX2:=MatrixExponential(X/2):\\
EY:=MatrixExponential(Y):\\
EXY2:=EX2.\!EY\!.EX2:\\
S:=MatrixFunction(EXY2, ln(x), x):\\
for n from 1 to N do: s[n]:= S[1,n+1]; od:
}

\smallskip
\noindent
Again, the most time consuming part of the process is writing the results to the screen.

\smallskip
\noindent
{\tt  
for n from 1 to N do: s[n]; od;
}

\smallskip
\noindent
Further variations on this theme are straightforward, and the relevant code will not be reproduced here.

\clearpage

\addcontentsline{toc}{section}{{\bf 
--------------------------------------------------------------------------------------------------------------------------}}
\section*{\bf Supplementary material}
\addcontentsline{toc}{section}{{\bf 
--------------------------------------------------------------------------------------------------------------------------}}
\begin{center}
\bf\Large Supplementary material for:\\ Simplifying the Reinsch algorithm for the  \\Baker--Campbell--Hausdorff series
\end{center}
\begin{quotation}
The Baker--Campbell--Hausdorff series computes the quantity 
\begin{equation*}
Z(X,Y)=\ln\left( e^X e^Y \right) = \sum_w g(w) \;  w(X,Y) =   \sum_{n=1}^\infty\left\{ \sum_{|w|=n} g(w)\; w(X,Y) \right\} = \sum_{n=1}^\infty z_n(X,Y), 
\end{equation*}
where $X$ and $Y$ are not necessarily commuting, in terms of homogeneous multinomials $z_n(X,Y)$ of degree $n$. In this supplementary material we report  explicit results for $z_5$, $z_6$,  $z_7$, and $z_8$, quantities which are still reasonably tractable.

\noindent
{\sc Keywords:}
Commutators, matrix exponentials, matrix logarithms, Baker--Campbell--Hausdorff formula.
\end{quotation}
\hrule
\section{Introduction}
\label{S:supplementary}

In this supplementary material to reference~\cite{simplified} we reproduce explicit results for $z_5$, $z_6$,  $z_7$, and $z_8$ in the BCH expansion, quantities which are still reasonably tractable.  Beyond this stage the formulae are simply too cumbersome to be usefully written down on paper. The only scientific justification for explicitly presenting even this level of detail is that it explicitly demonstrates the patterns and symmetries of the Goldberg coefficients $g(w)$ in a somewhat non-trivial context. These patterns and symmetries were useful in guiding our intuition when investigating  general properties of the Goldberg coefficients~\cite{simplified} , and in the subsequent analysis in which we develop bounds on the number of the non-zero Goldberg coefficients~\cite{simplified} .

One reason for going out as far as $z_8$, \emph{and no further},  is that the expression for $z_8$ is actually slightly shorter (by 2 terms) than $z_7$; while in contrast $z_9$ is significantly longer, (slightly over 3 times longer), and can no longer fit on a single page.  Another reason for going out to $z_8$, \emph{and no further}, is that $z_8$ contains the first Goldberg coefficient that is intrinsically rational, \begin{equation}
g(X^4Y^4)=g(Y^4X^4) = {23\over120960};
\end{equation}
all other Goldberg coefficients out to 8$^{th}$ order are either zero or reciprocals of integers.

\section{The $z_5$ term}

For $z_5$ our simplified algorithm~\cite{simplified}  yields:
\begin{eqnarray}
z_5 &=&
-{\frac {y_{{1}}y_{{2}}x_{{3}}y_{{4}}x_{{5}}}{120}}
+\frac{\,x_{{1}}y_{{2}}x_{{3}}y_{{4}}x_{{5}}}{30}
+\frac{\,y_{{1}}x_{{2}}y_{{3}}x_{{4}}y_{{5}}}{30}
-{\frac {y_{{1}}x_{{2}}y_{{3}}x_{{4}}x_{{5}}}{120}}
-{\frac {y_{{1}}x_{{2}}y_{{3}}y_{{4}}x_{{5}}}{120}}
-{\frac {y_{{1}}x_{{2}}x_{{3}}y_{{4}}x_{{5}}}{120}}
\nonumber\\
&&
-{\frac {y_{{1}}y_{{2}}y_{{3}}y_{{4}}x_{{5}}}{720}}
+{\frac {x_{{1}}y_{{2}}y_{{3}}y_{{4}}x_{{5}}}{180}}
-{\frac {x_{{1}}x_{{2}}y_{{3}}y_{{4}}x_{{5}}}{120}}
+{\frac {x_{{1}}x_{{2}}x_{{3}}y_{{4}}x_{{5}}}{180}}
+{\frac {y_{{1}}x_{{2}}y_{{3}}y_{{4}}y_{{5}}}{180}}
-{\frac {y_{{1}}x_{{2}}x_{{3}}y_{{4}}y_{{5}}}{120}}
\nonumber\\
&&
+{\frac {y_{{1}}x_{{2}}x_{{3}}x_{{4}}y_{{5}}}{180}}
-{\frac {y_{{1}}x_{{2}}x_{{3}}x_{{4}}x_{{5}}}{720}}
-{\frac {y_{{1}}y_{{2}}x_{{3}}y_{{4}}y_{{5}}}{120}}
-{\frac {y_{{1}}y_{{2}}x_{{3}}x_{{4}}y_{{5}}}{120}}
+{\frac {y_{{1}}y_{{2}}x_{{3}}x_{{4}}x_{{5}}}{180}}
-{\frac {x_{{1}}y_{{2}}x_{{3}}y_{{4}}y_{{5}}}{120}}
\nonumber\\
&&
-{\frac {x_{{1}}y_{{2}}x_{{3}}x_{{4}}y_{{5}}}{120}}
+{\frac {x_{{1}}y_{{2}}x_{{3}}x_{{4}}x_{{5}}}{180}}
+{\frac {y_{{1}}y_{{2}}y_{{3}}x_{{4}}y_{{5}}}{180}}
+{\frac {y_{{1}}y_{{2}}y_{{3}}x_{{4}}x_{{5}}}{180}}
-{\frac {x_{{1}}y_{{2}}y_{{3}}x_{{4}}y_{{5}}}{120}}
-{\frac {x_{{1}}y_{{2}}y_{{3}}x_{{4}}x_{{5}}}{120}}
\nonumber\\
&&
-{\frac {x_{{1}}x_{{2}}y_{{3}}x_{{4}}y_{{5}}}{120}}
-{\frac {x_{{1}}x_{{2}}y_{{3}}x_{{4}}x_{{5}}}{120}}
-{\frac {x_{{1}}y_{{2}}y_{{3}}y_{{4}}y_{{5}}}{720}}
+{\frac {x_{{1}}x_{{2}}y_{{3}}y_{{4}}y_{{5}}}{180}}
+{\frac {x_{{1}}x_{{2}}x_{{3}}y_{{4}}y_{{5}}}{180}}
-{\frac {x_{{1}}x_{{2}}x_{{3}}x_{{4}}y_{{5}}}{720}}.
\end{eqnarray}
Converting $x_i\to X$, and $y_i \to Y$ so as to get a representation in terms of words in the $\{X,Y\}$ alphabet, we obtain:
\begin{eqnarray}
z_5(X,Y) &=&
-{\frac {Y^2XYX}{120}}
+\frac{XYXYX}{30}
+\frac{YXYXY}{30}
-{\frac {YXYX^2}{120}}
-{\frac {YXY^2X}{120}}
-{\frac {YX^2YX}{120}}
\nonumber\\
&&
-{\frac {Y^4X}{720}}
+{\frac {XY^3X}{180}}
-{\frac {X^2Y^2X}{120}}
+{\frac {X^3 Y X}{180}}
+{\frac {YXY^3}{180}}
-{\frac {YX^2Y^2}{120}}
\nonumber\\
&&
+{\frac {YX^3Y}{180}}
-{\frac {YX^4}{720}}
-{\frac {Y^2XY^2}{120}}
-{\frac {Y^2X^2Y}{120}}
+{\frac {Y^2X^3}{180}}
-{\frac {XYXY^2}{120}}
\nonumber\\
&&
-{\frac {XYX^2Y}{120}}
+{\frac {XYX^3}{180}}
+{\frac {Y^3XY}{180}}
+{\frac {Y^3 X^2}{180}}
-{\frac {XY^2XY}{120}}
-{\frac {XY^2X^2}{120}}
\nonumber\\
&&
-{\frac {X^2YXY}{120}}
-{\frac {X^2YX^2}{120}}
-{\frac {XY^4}{720}}
+{\frac {X^2Y^3}{180}}
+{\frac {X^3 Y^2}{180}}
-{\frac {X^4 Y}{720}}.
\end{eqnarray}
Converting to right-nested commutators to get a Dynkin representation is tedious (due to various anti-symmetries and the Jacobi identity). For instance, Blanes and Casas~\cite{Blanes-Casas} report a commutator form of $z_5$ written in the Hall--Viennot basis. After a little work one can convert this into the equivalent right-nested long-commutator form:
\begin{equation}
z_5(X,Y) = {1\over6!}\left( - [X^4Y] + 6[XYXYX] +2 [XY^3X] +2 [YX^3Y] + 6[YXYXY] - [Y^4X]\right).
\end{equation}



\section{The $z_6$ term}

For $z_6$ our simplified algorithm~\cite{simplified}  yields:
\begin{eqnarray}
z_6 &=& 
-{\frac {y_{{1}}x_{{2}}y_{{3}}x_{{4}}y_{{5}}x_{{6}}}{60}}
-{\frac {y_{{1}}y_{{2}}y_{{3}}x_{{4}}y_{{5}}x_{{6}}}{360}}
+{\frac {y_{{1}}y_{{2}}x_{{3}}y_{{4}}x_{{5}}x_{{6}}}{240}}
+{\frac {y_{{1}}y_{{2}}x_{{3}}y_{{4}}y_{{5}}x_{{6}}}{240}}
+{\frac {y_{{1}}y_{{2}}x_{{3}}x_{{4}}y_{{5}}x_{{6}}}{240}}
+{\frac {x_{{1}}y_{{2}}x_{{3}}y_{{4}}x_{{5}}y_{{6}}}{60}}
\nonumber\\
&&
-{\frac {y_{{1}}x_{{2}}y_{{3}}x_{{4}}x_{{5}}x_{{6}}}{360}}
+{\frac {y_{{1}}x_{{2}}y_{{3}}y_{{4}}x_{{5}}x_{{6}}}{240}}
+{\frac {y_{{1}}x_{{2}}x_{{3}}y_{{4}}x_{{5}}x_{{6}}}{240}}
-{\frac {y_{{1}}x_{{2}}y_{{3}}y_{{4}}y_{{5}}x_{{6}}}{360}}
+{\frac {y_{{1}}x_{{2}}x_{{3}}y_{{4}}y_{{5}}x_{{6}}}{240}}
-{\frac {y_{{1}}x_{{2}}x_{{3}}x_{{4}}y_{{5}}x_{{6}}}{360}}
\nonumber\\
&&
-{\frac {x_{{1}}y_{{2}}y_{{3}}x_{{4}}y_{{5}}y_{{6}}}{240}}
-{\frac {x_{{1}}y_{{2}}y_{{3}}x_{{4}}x_{{5}}y_{{6}}}{240}}
-{\frac {x_{{1}}x_{{2}}y_{{3}}x_{{4}}y_{{5}}y_{{6}}}{240}}
-{\frac {x_{{1}}x_{{2}}y_{{3}}x_{{4}}x_{{5}}y_{{6}}}{240}}
+{\frac {y_{{1}}y_{{2}}y_{{3}}y_{{4}}x_{{5}}x_{{6}}}{1440}}
+{\frac {x_{{1}}y_{{2}}y_{{3}}y_{{4}}x_{{5}}y_{{6}}}{360}}
\nonumber\\
&&
-{\frac {x_{{1}}x_{{2}}y_{{3}}y_{{4}}x_{{5}}y_{{6}}}{240}}
+{\frac {x_{{1}}x_{{2}}x_{{3}}y_{{4}}x_{{5}}y_{{6}}}{360}}
+{\frac {y_{{1}}y_{{2}}x_{{3}}x_{{4}}x_{{5}}x_{{6}}}{1440}}
+{\frac {x_{{1}}y_{{2}}x_{{3}}y_{{4}}y_{{5}}y_{{6}}}{360}}
-{\frac {x_{{1}}y_{{2}}x_{{3}}x_{{4}}y_{{5}}y_{{6}}}{240}}
+{\frac {x_{{1}}y_{{2}}x_{{3}}x_{{4}}x_{{5}}y_{{6}}}{360}}
\nonumber\\
&&
-{\frac {y_{{1}}y_{{2}}y_{{3}}x_{{4}}x_{{5}}x_{{6}}}{360}}
+{\frac {x_{{1}}x_{{2}}x_{{3}}y_{{4}}y_{{5}}y_{{6}}}{360}}
-{\frac {x_{{1}}x_{{2}}x_{{3}}x_{{4}}y_{{5}}y_{{6}}}{1440}}
-{\frac {x_{{1}}x_{{2}}y_{{3}}y_{{4}}y_{{5}}y_{{6}}}{1440}}.
\end{eqnarray}
Converting $x_i\to X$ and $y_i \to Y$, so as to get a representation in terms of words in the $\{X,Y\}$ alphabet, we obtain:
\begin{eqnarray}
z_6(X,Y) &=& 
-{\frac {YXYXYX}{60}}
-{\frac {Y^3XYX}{360}}
+{\frac {Y^2XYX^2}{240}}
+{\frac {Y^2XY^2X}{240}}
+{\frac {Y^2X^2YX}{240}}
+{\frac {XYXYXY}{60}}
\nonumber\\
&&
-{\frac {YXYX^3}{360}}
+{\frac {YXY^2X^2}{240}}
+{\frac {YX^2YX^2}{240}}
-{\frac {YXY^3X}{360}}
+{\frac {YX^2Y^2X}{240}}
-{\frac {YX^3YX}{360}}
\nonumber\\
&&
-{\frac {XY^2XY^2}{240}}
-{\frac {XY^2X^2Y}{240}}
-{\frac {X^2YXY^2}{240}}
-{\frac {X^2YX^2Y}{240}}
+{\frac {Y^4X^2}{1440}}
+{\frac {XY^3XY}{360}}
\nonumber\\
&&-{
\frac {X^2Y^2XY}{240}}
+{\frac {X^3YXY}{360}}
+{\frac {Y^2X^4}{1440}}
+{\frac {XYXY^3}{360}}
-{\frac {XYX^2Y^2}{240}}
+{\frac {XYX^3Y}{360}}
\nonumber\\
&&
-{
\frac {Y^3X^3}{360}}
+{\frac {X^3Y^3}{360}}
-{\frac {X^4Y^2}{1440}}
-{\frac {X^2Y^4}{1440}}.
\end{eqnarray}
Converting to right-nested commutators to get a Dynkin representation is tedious (due to various anti-symmetries and the Jacobi identity). 
For instance, Blanes and Casas~\cite{Blanes-Casas} report a commutator form of $z_6$ written in the Hall--Viennot basis. After a little work one can convert this into the equivalent right-nested long-commutator form:
\begin{equation}
z_6(X,Y) = {1\over2\cdot6!}\left( - 2[X^2Y^2XY] + 6[XYXYXY] - [XY^4X] + [YX^4Y] \right).
\end{equation}

\vfill

\section{The $z_7$ term}
For $z_7$ our simplified algorithm~\cite{simplified}  yields:
\begin{eqnarray}
z_7 &=&
-{\frac {x_{{1}}y_{{2}}x_{{3}}y_{{4}}x_{{5}}y_{{6}}x_{{7}}}{140}}+{
\frac {y_{{1}}y_{{2}}x_{{3}}y_{{4}}x_{{5}}y_{{6}}x_{{7}}}{840}}+{
\frac {y_{{1}}x_{{2}}y_{{3}}y_{{4}}x_{{5}}y_{{6}}x_{{7}}}{840}}+{
\frac {y_{{1}}x_{{2}}x_{{3}}y_{{4}}x_{{5}}y_{{6}}x_{{7}}}{840}}-{
\frac {y_{{1}}x_{{2}}y_{{3}}x_{{4}}y_{{5}}x_{{6}}y_{{7}}}{140}}
\nonumber\\
&&
+{
\frac {y_{{1}}x_{{2}}y_{{3}}x_{{4}}y_{{5}}x_{{6}}x_{{7}}}{840}}+{
\frac {y_{{1}}x_{{2}}y_{{3}}x_{{4}}y_{{5}}y_{{6}}x_{{7}}}{840}}+{
\frac {y_{{1}}x_{{2}}y_{{3}}x_{{4}}x_{{5}}y_{{6}}x_{{7}}}{840}}-{
\frac {x_{{1}}x_{{2}}x_{{3}}y_{{4}}x_{{5}}y_{{6}}x_{{7}}}{630}}+{
\frac {y_{{1}}y_{{2}}y_{{3}}y_{{4}}x_{{5}}y_{{6}}x_{{7}}}{2016}}
\nonumber\\
&&
-{
\frac {x_{{1}}y_{{2}}y_{{3}}y_{{4}}x_{{5}}y_{{6}}x_{{7}}}{630}}+{
\frac {x_{{1}}x_{{2}}y_{{3}}y_{{4}}x_{{5}}y_{{6}}x_{{7}}}{840}}+{
\frac {x_{{1}}y_{{2}}y_{{3}}x_{{4}}y_{{5}}x_{{6}}y_{{7}}}{840}}+{
\frac {x_{{1}}y_{{2}}y_{{3}}x_{{4}}y_{{5}}x_{{6}}x_{{7}}}{840}}+{
\frac {x_{{1}}y_{{2}}y_{{3}}x_{{4}}y_{{5}}y_{{6}}x_{{7}}}{840}}
\nonumber\\
&&
+{
\frac {x_{{1}}y_{{2}}y_{{3}}x_{{4}}x_{{5}}y_{{6}}x_{{7}}}{840}}+{
\frac {x_{{1}}x_{{2}}y_{{3}}x_{{4}}y_{{5}}x_{{6}}y_{{7}}}{840}}+{
\frac {x_{{1}}x_{{2}}y_{{3}}x_{{4}}y_{{5}}x_{{6}}x_{{7}}}{840}}+{
\frac {x_{{1}}x_{{2}}y_{{3}}x_{{4}}y_{{5}}y_{{6}}x_{{7}}}{840}}+{
\frac {x_{{1}}x_{{2}}y_{{3}}x_{{4}}x_{{5}}y_{{6}}x_{{7}}}{840}}
\nonumber\\
&&
-{
\frac {y_{{1}}y_{{2}}y_{{3}}x_{{4}}y_{{5}}x_{{6}}y_{{7}}}{630}}-{
\frac {y_{{1}}y_{{2}}y_{{3}}x_{{4}}y_{{5}}x_{{6}}x_{{7}}}{5040}}-{
\frac {y_{{1}}y_{{2}}y_{{3}}x_{{4}}y_{{5}}y_{{6}}x_{{7}}}{5040}}-{
\frac {y_{{1}}y_{{2}}y_{{3}}x_{{4}}x_{{5}}y_{{6}}x_{{7}}}{5040}}+{
\frac {x_{{1}}y_{{2}}x_{{3}}y_{{4}}x_{{5}}y_{{6}}y_{{7}}}{840}}
\nonumber\\
&&+{
\frac {x_{{1}}y_{{2}}x_{{3}}y_{{4}}x_{{5}}x_{{6}}y_{{7}}}{840}}-{
\frac {x_{{1}}y_{{2}}x_{{3}}y_{{4}}x_{{5}}x_{{6}}x_{{7}}}{630}}+{
\frac {x_{{1}}y_{{2}}x_{{3}}y_{{4}}y_{{5}}x_{{6}}y_{{7}}}{840}}+{
\frac {x_{{1}}y_{{2}}x_{{3}}y_{{4}}y_{{5}}x_{{6}}x_{{7}}}{840}}+{
\frac {x_{{1}}y_{{2}}x_{{3}}x_{{4}}y_{{5}}x_{{6}}y_{{7}}}{840}}
\nonumber\\
&&+{
\frac {x_{{1}}y_{{2}}x_{{3}}x_{{4}}y_{{5}}x_{{6}}x_{{7}}}{840}}-{
\frac {x_{{1}}y_{{2}}x_{{3}}y_{{4}}y_{{5}}y_{{6}}x_{{7}}}{630}}+{
\frac {x_{{1}}y_{{2}}x_{{3}}x_{{4}}y_{{5}}y_{{6}}x_{{7}}}{840}}-{
\frac {x_{{1}}y_{{2}}x_{{3}}x_{{4}}x_{{5}}y_{{6}}x_{{7}}}{630}}+{
\frac {y_{{1}}y_{{2}}x_{{3}}y_{{4}}x_{{5}}y_{{6}}y_{{7}}}{840}}
\nonumber\\
&&+{
\frac {y_{{1}}y_{{2}}x_{{3}}y_{{4}}x_{{5}}x_{{6}}y_{{7}}}{840}}-{
\frac {y_{{1}}y_{{2}}x_{{3}}y_{{4}}x_{{5}}x_{{6}}x_{{7}}}{5040}}+{
\frac {y_{{1}}y_{{2}}x_{{3}}y_{{4}}y_{{5}}x_{{6}}y_{{7}}}{840}}-{
\frac {y_{{1}}y_{{2}}x_{{3}}y_{{4}}y_{{5}}x_{{6}}x_{{7}}}{1120}}+{
\frac {y_{{1}}y_{{2}}x_{{3}}x_{{4}}y_{{5}}x_{{6}}y_{{7}}}{840}}
\nonumber\\
&&
-{
\frac {y_{{1}}y_{{2}}x_{{3}}x_{{4}}y_{{5}}x_{{6}}x_{{7}}}{1120}}-{
\frac {y_{{1}}y_{{2}}x_{{3}}y_{{4}}y_{{5}}y_{{6}}x_{{7}}}{5040}}-{
\frac {y_{{1}}y_{{2}}x_{{3}}x_{{4}}y_{{5}}y_{{6}}x_{{7}}}{1120}}-{
\frac {y_{{1}}y_{{2}}x_{{3}}x_{{4}}x_{{5}}y_{{6}}x_{{7}}}{5040}}-{
\frac {y_{{1}}x_{{2}}y_{{3}}x_{{4}}x_{{5}}x_{{6}}y_{{7}}}{630}}
\nonumber\\
&&
+{
\frac {y_{{1}}x_{{2}}y_{{3}}x_{{4}}x_{{5}}x_{{6}}x_{{7}}}{2016}}+{
\frac {y_{{1}}x_{{2}}y_{{3}}y_{{4}}x_{{5}}y_{{6}}y_{{7}}}{840}}+{
\frac {y_{{1}}x_{{2}}y_{{3}}y_{{4}}x_{{5}}x_{{6}}y_{{7}}}{840}}-{
\frac {y_{{1}}x_{{2}}y_{{3}}y_{{4}}x_{{5}}x_{{6}}x_{{7}}}{5040}}+{
\frac {y_{{1}}x_{{2}}x_{{3}}y_{{4}}x_{{5}}y_{{6}}y_{{7}}}{840}}
\nonumber\\
&&
+{
\frac {y_{{1}}x_{{2}}x_{{3}}y_{{4}}x_{{5}}x_{{6}}y_{{7}}}{840}}-{
\frac {y_{{1}}x_{{2}}x_{{3}}y_{{4}}x_{{5}}x_{{6}}x_{{7}}}{5040}}-{
\frac {y_{{1}}x_{{2}}y_{{3}}y_{{4}}y_{{5}}x_{{6}}y_{{7}}}{630}}-{
\frac {y_{{1}}x_{{2}}y_{{3}}y_{{4}}y_{{5}}x_{{6}}x_{{7}}}{5040}}+{
\frac {y_{{1}}x_{{2}}x_{{3}}y_{{4}}y_{{5}}x_{{6}}y_{{7}}}{840}}
\nonumber\\
&&-{
\frac {y_{{1}}x_{{2}}x_{{3}}y_{{4}}y_{{5}}x_{{6}}x_{{7}}}{1120}}
-{
\frac {y_{{1}}x_{{2}}x_{{3}}x_{{4}}y_{{5}}x_{{6}}y_{{7}}}{630}}-{
\frac {y_{{1}}x_{{2}}x_{{3}}x_{{4}}y_{{5}}x_{{6}}x_{{7}}}{5040}}+{
\frac {y_{{1}}x_{{2}}y_{{3}}y_{{4}}y_{{5}}y_{{6}}x_{{7}}}{2016}}-{
\frac {y_{{1}}x_{{2}}x_{{3}}y_{{4}}y_{{5}}y_{{6}}x_{{7}}}{5040}}
\nonumber\\
&&
-{
\frac {y_{{1}}x_{{2}}x_{{3}}x_{{4}}y_{{5}}y_{{6}}x_{{7}}}{5040}}+{
\frac {y_{{1}}x_{{2}}x_{{3}}x_{{4}}x_{{5}}y_{{6}}x_{{7}}}{2016}}-{
\frac {y_{{1}}x_{{2}}y_{{3}}x_{{4}}y_{{5}}y_{{6}}y_{{7}}}{630}}+{
\frac {y_{{1}}x_{{2}}y_{{3}}x_{{4}}x_{{5}}y_{{6}}y_{{7}}}{840}}-{
\frac {x_{{1}}y_{{2}}y_{{3}}y_{{4}}y_{{5}}y_{{6}}x_{{7}}}{5040}}
\nonumber\\
&&+{
\frac {x_{{1}}x_{{2}}y_{{3}}y_{{4}}y_{{5}}y_{{6}}x_{{7}}}{2016}}-{
\frac {x_{{1}}x_{{2}}x_{{3}}y_{{4}}y_{{5}}y_{{6}}x_{{7}}}{1512}}+{
\frac {x_{{1}}x_{{2}}x_{{3}}x_{{4}}y_{{5}}y_{{6}}x_{{7}}}{2016}}-{
\frac {x_{{1}}x_{{2}}x_{{3}}x_{{4}}x_{{5}}y_{{6}}x_{{7}}}{5040}}-{
\frac {x_{{1}}y_{{2}}y_{{3}}y_{{4}}x_{{5}}x_{{6}}x_{{7}}}{1512}}
\nonumber\\
&&-{
\frac {x_{{1}}x_{{2}}y_{{3}}y_{{4}}x_{{5}}y_{{6}}y_{{7}}}{1120}}-{
\frac {x_{{1}}x_{{2}}y_{{3}}y_{{4}}x_{{5}}x_{{6}}y_{{7}}}{1120}}-{
\frac {x_{{1}}x_{{2}}y_{{3}}y_{{4}}x_{{5}}x_{{6}}x_{{7}}}{5040}}-{
\frac {x_{{1}}x_{{2}}x_{{3}}y_{{4}}x_{{5}}y_{{6}}y_{{7}}}{5040}}-{
\frac {x_{{1}}x_{{2}}x_{{3}}y_{{4}}x_{{5}}x_{{6}}y_{{7}}}{5040}}
\nonumber\\
&&-{
\frac {x_{{1}}x_{{2}}x_{{3}}y_{{4}}x_{{5}}x_{{6}}x_{{7}}}{1512}}-{
\frac {y_{{1}}y_{{2}}y_{{3}}y_{{4}}y_{{5}}x_{{6}}y_{{7}}}{5040}}-{
\frac {y_{{1}}y_{{2}}y_{{3}}y_{{4}}y_{{5}}x_{{6}}x_{{7}}}{5040}}+{
\frac {x_{{1}}y_{{2}}y_{{3}}y_{{4}}y_{{5}}x_{{6}}y_{{7}}}{2016}}+{
\frac {x_{{1}}y_{{2}}y_{{3}}y_{{4}}y_{{5}}x_{{6}}x_{{7}}}{2016}}
\nonumber\\
&&-{
\frac {x_{{1}}x_{{2}}y_{{3}}y_{{4}}y_{{5}}x_{{6}}y_{{7}}}{5040}}-{
\frac {x_{{1}}x_{{2}}y_{{3}}y_{{4}}y_{{5}}x_{{6}}x_{{7}}}{5040}}-{
\frac {x_{{1}}x_{{2}}x_{{3}}y_{{4}}y_{{5}}x_{{6}}y_{{7}}}{5040}}-{
\frac {x_{{1}}x_{{2}}x_{{3}}y_{{4}}y_{{5}}x_{{6}}x_{{7}}}{5040}}+{
\frac {x_{{1}}x_{{2}}x_{{3}}x_{{4}}y_{{5}}x_{{6}}y_{{7}}}{2016}}
\nonumber\\
&&+{
\frac {x_{{1}}x_{{2}}x_{{3}}x_{{4}}y_{{5}}x_{{6}}x_{{7}}}{2016}}+{
\frac {y_{{1}}y_{{2}}y_{{3}}y_{{4}}y_{{5}}y_{{6}}x_{{7}}}{30240}}-{
\frac {x_{{1}}y_{{2}}x_{{3}}x_{{4}}x_{{5}}x_{{6}}x_{{7}}}{5040}}-{
\frac {y_{{1}}y_{{2}}y_{{3}}x_{{4}}y_{{5}}y_{{6}}y_{{7}}}{1512}}-{
\frac {y_{{1}}y_{{2}}y_{{3}}x_{{4}}x_{{5}}y_{{6}}y_{{7}}}{5040}}
\nonumber\\
&&-{
\frac {y_{{1}}y_{{2}}y_{{3}}x_{{4}}x_{{5}}x_{{6}}y_{{7}}}{1512}}+{
\frac {y_{{1}}y_{{2}}y_{{3}}x_{{4}}x_{{5}}x_{{6}}x_{{7}}}{3780}}-{
\frac {x_{{1}}y_{{2}}y_{{3}}x_{{4}}y_{{5}}y_{{6}}y_{{7}}}{5040}}-{
\frac {x_{{1}}y_{{2}}y_{{3}}x_{{4}}x_{{5}}y_{{6}}y_{{7}}}{1120}}-{
\frac {x_{{1}}y_{{2}}y_{{3}}x_{{4}}x_{{5}}x_{{6}}y_{{7}}}{5040}}
\nonumber\\
&&+{
\frac {x_{{1}}y_{{2}}y_{{3}}x_{{4}}x_{{5}}x_{{6}}x_{{7}}}{2016}}-{
\frac {x_{{1}}x_{{2}}y_{{3}}x_{{4}}y_{{5}}y_{{6}}y_{{7}}}{5040}}-{
\frac {x_{{1}}x_{{2}}y_{{3}}x_{{4}}x_{{5}}y_{{6}}y_{{7}}}{1120}}-{
\frac {x_{{1}}x_{{2}}y_{{3}}x_{{4}}x_{{5}}x_{{6}}y_{{7}}}{5040}}+{
\frac {x_{{1}}x_{{2}}y_{{3}}x_{{4}}x_{{5}}x_{{6}}x_{{7}}}{2016}}
\nonumber\\
&&+{
\frac {y_{{1}}y_{{2}}y_{{3}}y_{{4}}x_{{5}}y_{{6}}y_{{7}}}{2016}}+{
\frac {y_{{1}}y_{{2}}y_{{3}}y_{{4}}x_{{5}}x_{{6}}y_{{7}}}{2016}}+{
\frac {y_{{1}}y_{{2}}y_{{3}}y_{{4}}x_{{5}}x_{{6}}x_{{7}}}{3780}}-{
\frac {x_{{1}}y_{{2}}y_{{3}}y_{{4}}x_{{5}}y_{{6}}y_{{7}}}{5040}}-{
\frac {x_{{1}}y_{{2}}y_{{3}}y_{{4}}x_{{5}}x_{{6}}y_{{7}}}{5040}}
\nonumber\\
&&-{
\frac {y_{{1}}x_{{2}}y_{{3}}y_{{4}}y_{{5}}y_{{6}}y_{{7}}}{5040}}+{
\frac {y_{{1}}x_{{2}}x_{{3}}y_{{4}}y_{{5}}y_{{6}}y_{{7}}}{2016}}-{
\frac {y_{{1}}x_{{2}}x_{{3}}x_{{4}}y_{{5}}y_{{6}}y_{{7}}}{1512}}+{
\frac {y_{{1}}x_{{2}}x_{{3}}x_{{4}}x_{{5}}y_{{6}}y_{{7}}}{2016}}-{
\frac {y_{{1}}x_{{2}}x_{{3}}x_{{4}}x_{{5}}x_{{6}}y_{{7}}}{5040}}
\nonumber\\
&&+{
\frac {y_{{1}}x_{{2}}x_{{3}}x_{{4}}x_{{5}}x_{{6}}x_{{7}}}{30240}}+{
\frac {y_{{1}}y_{{2}}x_{{3}}y_{{4}}y_{{5}}y_{{6}}y_{{7}}}{2016}}-{
\frac {y_{{1}}y_{{2}}x_{{3}}x_{{4}}y_{{5}}y_{{6}}y_{{7}}}{5040}}-{
\frac {y_{{1}}y_{{2}}x_{{3}}x_{{4}}x_{{5}}y_{{6}}y_{{7}}}{5040}}+{
\frac {y_{{1}}y_{{2}}x_{{3}}x_{{4}}x_{{5}}x_{{6}}y_{{7}}}{2016}}
\nonumber\\
&&-{
\frac {y_{{1}}y_{{2}}x_{{3}}x_{{4}}x_{{5}}x_{{6}}x_{{7}}}{5040}}+{
\frac {x_{{1}}y_{{2}}x_{{3}}y_{{4}}y_{{5}}y_{{6}}y_{{7}}}{2016}}-{
\frac {x_{{1}}y_{{2}}x_{{3}}x_{{4}}y_{{5}}y_{{6}}y_{{7}}}{5040}}-{
\frac {x_{{1}}y_{{2}}x_{{3}}x_{{4}}x_{{5}}y_{{6}}y_{{7}}}{5040}}+{
\frac {x_{{1}}y_{{2}}x_{{3}}x_{{4}}x_{{5}}x_{{6}}y_{{7}}}{2016}}
\nonumber\\
&&+{
\frac {x_{{1}}y_{{2}}y_{{3}}y_{{4}}y_{{5}}y_{{6}}y_{{7}}}{30240}}-{
\frac {x_{{1}}x_{{2}}y_{{3}}y_{{4}}y_{{5}}y_{{6}}y_{{7}}}{5040}}+{
\frac {x_{{1}}x_{{2}}x_{{3}}y_{{4}}y_{{5}}y_{{6}}y_{{7}}}{3780}}+{
\frac {x_{{1}}x_{{2}}x_{{3}}x_{{4}}y_{{5}}y_{{6}}y_{{7}}}{3780}}-{
\frac {x_{{1}}x_{{2}}x_{{3}}x_{{4}}x_{{5}}y_{{6}}y_{{7}}}{5040}}
\nonumber\\
&&+{
\frac {x_{{1}}x_{{2}}x_{{3}}x_{{4}}x_{{5}}x_{{6}}y_{{7}}}{30240}}.
\end{eqnarray}
There are 63 occurrences of each $x_i$ above, and 63 occurrences of each $y_i$.
(Note that $63+63=126 = \hbox{number of terms}$.)
Furthermore the coefficients appearing above are either zero or reciprocals of integers. 

\clearpage

Converting $x_i\to X$ and $y_i \to Y$ so as to get a representation in terms of words in the $\{X,Y\}$ alphabet we obtain:
 {\small
\begin{eqnarray}
{\large z_7(X,Y) }&=&
-{\frac {XYXYXYX}{140}}
+{\frac {Y^2XYXYX}{840}}
+{\frac {YXY^2XYX}{840}}
+{\frac {YX^2YXYX}{840}}
-{\frac {YXYXYXY}{140}}
\nonumber\\
&&
+{\frac {YXYXYX^2}{840}}
+{\frac {YXYXY^2X}{840}}
+{\frac {YXYX^2YX}{840}}
-{\frac {X^3YXYX}{630}}
+{\frac {Y^4XYX }{2016}}
\nonumber\\
&&
-{\frac {XY^3XYX}{630}}
+{\frac {X^2Y^2XYX}{840}}
+{\frac {XY^2XYXY}{840}}
+{\frac {XY^2XYX^2}{840}}
+{\frac {XY^2XY^2X}{840}}
\nonumber\\
&&
+{\frac {XY^2X^2YX}{840}}
+{\frac {X^2YXYXY}{840}}
+{\frac {X^2YXYX^2 }{840}}
+{\frac {X^2YXY^2X}{840}}
+{\frac {X^2YX^2YX}{840}}
\nonumber\\
&&
-{\frac {Y^3XYXY }{630}}
-{\frac {Y^3XYX^2}{5040}}
-{\frac {Y^3XY^2X}{5040}}
-{\frac {Y^3X^2YX}{5040}}
+{\frac {XYXYXY^2}{840}}
\nonumber\\
&&
+{\frac {XYXYX^2Y}{840}}
-{\frac {XYXYX^3}{630}}
+{\frac {XYXY^2XY}{840}}
+{\frac {XYXY^2X^2}{840}}
+{\frac { XYX^2YXY}{840}}
\nonumber\\
&&
+{\frac {XYX^2YX^2}{840}}
-{\frac {XYXY^3X}{630}}
+{\frac {XYX^2Y^2X}{840}}
-{\frac {XYX^3YX}{630}}
+{\frac {Y^2XYXY^2}{840}}
\nonumber\\
&&
+{
\frac {Y^2XYX^2Y}{840}}-{
\frac {Y^2XYX^3}{5040}}+{
\frac {Y^2XY^2XY}{840}}-{
\frac {Y^2XY^2X^2}{1120}}+{
\frac {Y^2X^2YXY}{840}}
\nonumber\\
&&
-{
\frac {Y^2X^2YX^2}{1120}}-{
\frac {Y^2XY^3X}{5040}}-{
\frac {Y^2X^2Y^2X}{1120}}-{
\frac {Y^2X^3YX}{5040}}-{
\frac {YXYX^3Y}{630}}
\nonumber\\
&&
+{
\frac {YXYX^4}{2016}}+{
\frac {YXY^2XY^2}{840}}+{
\frac {YXY^2X^2Y}{840}}-{
\frac {YXY^2X^3}{5040}}+{
\frac {YX^2YXY^2}{840}}
\nonumber\\
&&
+{
\frac {YX^2YX^2Y}{840}}-{
\frac {YX^2YX^3}{5040}}-{
\frac {YXY^3XY}{630}}-{
\frac {YXY^3X^2}{5040}}+{
\frac {YX^2Y^2XY}{840}}
\nonumber\\
&&
-{\frac {YX^2Y^2X^2}{1120}}
-{\frac {YX^3YXY}{630}}
-{\frac {YX^3YX^2}{5040}}
+{\frac {YXY^4X}{2016}}
-{\frac {YX^2Y^3 X}{5040}}
\nonumber\\
&&
-{
\frac {YX^3Y^2X}{5040}}+{
\frac {YX^4YX}{2016}}-{
\frac {YXYXY^3}{630}}+{
\frac {YXYX^2Y^2}{840}}-{
\frac {XY^5X}{5040}}
\nonumber\\
&&
+{
\frac {X^2Y^4X}{2016}}-{
\frac {X^3Y^3X}{1512}}+{
\frac {X^4Y^2X}{2016}}-{
\frac {X^5YX}{5040}}-{
\frac {XY^3X^3}{1512}}
\nonumber\\
&&
-{
\frac {X^2Y^2XY^2}{1120}}-{
\frac {X^2Y^2X^2Y}{1120}}-{
\frac {X^2Y^2X^3}{5040}}-{
\frac {X^3YXY^2}{5040}}-{
\frac {X^3YX^2Y}{5040}}
\nonumber\\
&&
-{
\frac {X^3YX^3}{1512}}-{
\frac {Y^5XY}{5040}}-{
\frac {Y^5X^2}{5040}}+{
\frac {XY^4XY}{2016}}+{
\frac {XY^4X^2}{2016}}
\nonumber\\
&&
-{
\frac {X^2Y^3XY}{5040}}-{
\frac {X^2Y^3X^2}{5040}}-{
\frac {X^3Y^2XY}{5040}}-{
\frac {X^3Y^2X^2}{5040}}+{
\frac {X^4YXY}{2016}}
\nonumber\\
&&
+{\frac {X^4YX^2}{2016}}
+{\frac {Y^6 X }{30240}}
-{\frac {XYX^5}{5040}}
-{\frac {Y^3XY^3}{1512}}
-{\frac {Y^3X^2Y^2}{5040}}
\nonumber\\
&&
-{\frac {Y^3X^3Y}{1512}}
+{\frac {Y^3X^4}{3780}}
-{\frac {XY^2XY^3}{5040}}
-{\frac {XY^2X^2Y^2}{1120}}
-{\frac {XY^2X^3Y}{5040}}
\nonumber\\
&&
+{
\frac {XY^2X^4}{2016}}-{
\frac {X^2YXY^3}{5040}}-{
\frac {X^2YX^2Y^2}{1120}}-{
\frac {X^2YX^3Y}{5040}}+{
\frac {X^2YX^4}{2016}}
\nonumber\\
&&
+{
\frac {Y^4XY^2}{2016}}+{
\frac {Y^4X^2Y}{2016}}+{
\frac {Y^4X^3}{3780}}-{
\frac {XY^3 XY^2}{5040}}-{
\frac {XY^3X^2Y  }{5040}}
\nonumber\\
&&
-{
\frac {YXY^5}{5040}}+{
\frac {YX^2Y^4}{2016}}-{
\frac {YX^3Y^3 }{1512}}+{
\frac {YX^4Y^2}{2016}}-{
\frac {YX^5Y}{5040}}
\nonumber\\
&&
+{
\frac {YX^6}{30240}}+{
\frac {Y^2XY^4}{2016}}-{
\frac {Y^2X^2Y^3}{5040}}-{
\frac {Y^2X^3Y^2}{5040}}+{
\frac {Y^2X^4Y }{2016}}
\nonumber\\
&&-{
\frac {Y^2X^5}{5040}}+{
\frac {XYXY^4}{2016}}-{
\frac {XYX^2Y^3}{5040}}-{
\frac {XYX^3Y^2 }{5040}}+{
\frac {XYX^4Y}{2016}}
\nonumber\\
&&
+{
\frac {XY^6}{30240}}-{
\frac {X^2Y^5}{5040}}+{
\frac {X^3Y^4}{3780}}+{
\frac {X^4Y^3}{3780}}-{
\frac {X^5Y^2}{5040}}
\nonumber\\
&&+{
\frac {X^6Y}{30240}}.
\end{eqnarray}}
\enlargethispage{35pt}\!
There are 2 occurrences of $X^6$ above,
5 occurrences of $X^5$,
12 occurrences of $X^4$,
28 occurrences of $X^3$,
64 occurrences of $X^2$,
and 144 occurrences of $X^1$.
Similarly for $Y$.
(Note that $63\times 7 = 2\times 6 + 5\times5+12\times4+28\times3+68\times2+144\times 1$.)
Converting this to commutators, (in any form, either the Dynkin form or using other commutator bases), is impractical without significant additional computer-aided computation. See for instance reference~\cite{Casas-Murua}, and related online tables~\cite{Murua-tables}. 

\section{The $z_8$ term}
\noindent
For $z_8$ our simplified algorithm~\cite{simplified}  yields:
\begin{eqnarray}
z_8 &=&
{\frac {y_{{1}}y_{{2}}y_{{3}}x_{{4}}y_{{5}}x_{{6}}y_{{7}}x_{{8}}}{1260}}-
{\frac {x_{{1}}y_{{2}}x_{{3}}y_{{4}}x_{{5}}y_{{6}}x_{{7}}y_{{8}}}{280}}-
{\frac {y_{{1}}y_{{2}}x_{{3}}y_{{4}}y_{{5}}x_{{6}}y_{{7}}x_{{8}}}{1680}}-
{\frac {y_{{1}}y_{{2}}x_{{3}}x_{{4}}y_{{5}}x_{{6}}y_{{7}}x_{{8}}}{1680}}-
{\frac {y_{{1}}y_{{2}}x_{{3}}y_{{4}}x_{{5}}y_{{6}}x_{{7}}x_{{8}}}{1680}}
\nonumber\\
&&
-{\frac {y_{{1}}y_{{2}}x_{{3}}y_{{4}}x_{{5}}y_{{6}}y_{{7}}x_{{8}}}{1680}}-
{\frac {y_{{1}}y_{{2}}x_{{3}}y_{{4}}x_{{5}}x_{{6}}y_{{7}}x_{{8}}}{1680}}-
{\frac {y_{{1}}x_{{2}}y_{{3}}y_{{4}}x_{{5}}y_{{6}}x_{{7}}x_{{8}}}{1680}}-
{\frac {y_{{1}}x_{{2}}y_{{3}}y_{{4}}x_{{5}}y_{{6}}y_{{7}}x_{{8}}}{1680}}-
{\frac {y_{{1}}x_{{2}}y_{{3}}y_{{4}}x_{{5}}x_{{6}}y_{{7}}x_{{8}}}{1680}}
\nonumber\\
&&
-{\frac {y_{{1}}x_{{2}}y_{{3}}x_{{4}}x_{{5}}y_{{6}}x_{{7}}x_{{8}}}{1680}}+
{\frac {y_{{1}}x_{{2}}y_{{3}}x_{{4}}y_{{5}}y_{{6}}y_{{7}}x_{{8}}}{1260}}-
{\frac {y_{{1}}x_{{2}}y_{{3}}x_{{4}}x_{{5}}y_{{6}}y_{{7}}x_{{8}}}{1680}}+
{\frac {y_{{1}}x_{{2}}y_{{3}}x_{{4}}x_{{5}}x_{{6}}y_{{7}}x_{{8}}}{1260}}+
{\frac {y_{{1}}x_{{2}}y_{{3}}x_{{4}}y_{{5}}x_{{6}}x_{{7}}x_{{8}}}{1260}}
\nonumber\\
&&
-{\frac {y_{{1}}x_{{2}}y_{{3}}x_{{4}}y_{{5}}y_{{6}}x_{{7}}x_{{8}}}{1680}}+
{\frac {y_{{1}}x_{{2}}y_{{3}}y_{{4}}y_{{5}}x_{{6}}y_{{7}}x_{{8}}}{1260}}-
{\frac {y_{{1}}x_{{2}}x_{{3}}y_{{4}}y_{{5}}x_{{6}}y_{{7}}x_{{8}}}{1680}}+
{\frac {y_{{1}}x_{{2}}x_{{3}}x_{{4}}y_{{5}}x_{{6}}y_{{7}}x_{{8}}}{1260}}-
{\frac {y_{{1}}x_{{2}}x_{{3}}y_{{4}}x_{{5}}y_{{6}}x_{{7}}x_{{8}}}{1680}}
\nonumber\\
&&
-{\frac {y_{{1}}x_{{2}}x_{{3}}y_{{4}}x_{{5}}y_{{6}}y_{{7}}x_{{8}}}{1680}}-
{\frac {y_{{1}}x_{{2}}x_{{3}}y_{{4}}x_{{5}}x_{{6}}y_{{7}}x_{{8}}}{1680}}+
{\frac {y_{{1}}y_{{2}}y_{{3}}y_{{4}}y_{{5}}x_{{6}}y_{{7}}x_{{8}}}{10080}}+
{\frac {x_{{1}}x_{{2}}y_{{3}}y_{{4}}x_{{5}}y_{{6}}x_{{7}}y_{{8}}}{1680}}-
{\frac {x_{{1}}x_{{2}}x_{{3}}y_{{4}}x_{{5}}y_{{6}}x_{{7}}y_{{8}}}{1260}}
\nonumber\\
&&
-{\frac {y_{{1}}y_{{2}}y_{{3}}y_{{4}}x_{{5}}y_{{6}}x_{{7}}x_{{8}}}{4032}}-
{\frac {y_{{1}}y_{{2}}y_{{3}}y_{{4}}x_{{5}}y_{{6}}y_{{7}}x_{{8}}}{4032}}-
{\frac {y_{{1}}y_{{2}}y_{{3}}y_{{4}}x_{{5}}x_{{6}}y_{{7}}x_{{8}}}{4032}}-
{\frac {x_{{1}}y_{{2}}y_{{3}}y_{{4}}x_{{5}}y_{{6}}x_{{7}}y_{{8}}}{1260}}+
{\frac {x_{{1}}x_{{2}}y_{{3}}x_{{4}}x_{{5}}y_{{6}}x_{{7}}y_{{8}}}{1680}}
\nonumber\\
&&
+{\frac {x_{{1}}y_{{2}}y_{{3}}x_{{4}}y_{{5}}y_{{6}}x_{{7}}y_{{8}}}{1680}}+
{\frac {x_{{1}}y_{{2}}y_{{3}}x_{{4}}x_{{5}}y_{{6}}x_{{7}}y_{{8}}}{1680}}+
{\frac {x_{{1}}x_{{2}}y_{{3}}x_{{4}}y_{{5}}x_{{6}}y_{{7}}y_{{8}}}{1680}}+
{\frac {x_{{1}}x_{{2}}y_{{3}}x_{{4}}y_{{5}}x_{{6}}x_{{7}}y_{{8}}}{1680}}+
{\frac {x_{{1}}x_{{2}}y_{{3}}x_{{4}}y_{{5}}y_{{6}}x_{{7}}y_{{8}}}{1680}}
\nonumber\\
&&
+{\frac {y_{{1}}y_{{2}}y_{{3}}x_{{4}}y_{{5}}x_{{6}}x_{{7}}x_{{8}}}{3024}}+
{\frac {y_{{1}}y_{{2}}y_{{3}}x_{{4}}y_{{5}}y_{{6}}x_{{7}}x_{{8}}}{10080}}+
{\frac {y_{{1}}y_{{2}}y_{{3}}x_{{4}}x_{{5}}y_{{6}}x_{{7}}x_{{8}}}{10080}}+
{\frac {x_{{1}}y_{{2}}y_{{3}}x_{{4}}y_{{5}}x_{{6}}y_{{7}}y_{{8}}}{1680}}+
{\frac {x_{{1}}y_{{2}}y_{{3}}x_{{4}}y_{{5}}x_{{6}}x_{{7}}y_{{8}}}{1680}}
\nonumber\\
&&
+{\frac {y_{{1}}y_{{2}}y_{{3}}x_{{4}}y_{{5}}y_{{6}}y_{{7}}x_{{8}}}{3024}}+
{\frac {y_{{1}}y_{{2}}y_{{3}}x_{{4}}x_{{5}}y_{{6}}y_{{7}}x_{{8}}}{10080}}+
{\frac {y_{{1}}y_{{2}}y_{{3}}x_{{4}}x_{{5}}x_{{6}}y_{{7}}x_{{8}}}{3024}}-
{\frac {x_{{1}}y_{{2}}x_{{3}}y_{{4}}x_{{5}}y_{{6}}y_{{7}}y_{{8}}}{1260}}+
{\frac {x_{{1}}y_{{2}}x_{{3}}y_{{4}}x_{{5}}x_{{6}}y_{{7}}y_{{8}}}{1680}}
\nonumber\\
&&
-{\frac {x_{{1}}y_{{2}}x_{{3}}y_{{4}}x_{{5}}x_{{6}}x_{{7}}y_{{8}}}{1260}}+
{\frac {x_{{1}}y_{{2}}x_{{3}}y_{{4}}y_{{5}}x_{{6}}y_{{7}}y_{{8}}}{1680}}+
{\frac {x_{{1}}y_{{2}}x_{{3}}y_{{4}}y_{{5}}x_{{6}}x_{{7}}y_{{8}}}{1680}}+
{\frac {x_{{1}}y_{{2}}x_{{3}}x_{{4}}y_{{5}}x_{{6}}y_{{7}}y_{{8}}}{1680}}+
{\frac {x_{{1}}y_{{2}}x_{{3}}x_{{4}}y_{{5}}x_{{6}}x_{{7}}y_{{8}}}{1680}}
\nonumber\\
&&
-{\frac {x_{{1}}y_{{2}}x_{{3}}y_{{4}}y_{{5}}y_{{6}}x_{{7}}y_{{8}}}{1260}}+
{\frac {x_{{1}}y_{{2}}x_{{3}}x_{{4}}y_{{5}}y_{{6}}x_{{7}}y_{{8}}}{1680}}-
{\frac {x_{{1}}y_{{2}}x_{{3}}x_{{4}}x_{{5}}y_{{6}}x_{{7}}y_{{8}}}{1260}}+
{\frac {y_{{1}}y_{{2}}x_{{3}}y_{{4}}y_{{5}}x_{{6}}x_{{7}}x_{{8}}}{10080}}+
{\frac {y_{{1}}y_{{2}}x_{{3}}x_{{4}}y_{{5}}x_{{6}}x_{{7}}x_{{8}}}{10080}}
\nonumber\\
&&
+{\frac {y_{{1}}y_{{2}}x_{{3}}y_{{4}}y_{{5}}y_{{6}}x_{{7}}x_{{8}}}{10080}}+
{\frac {y_{{1}}y_{{2}}x_{{3}}x_{{4}}y_{{5}}y_{{6}}x_{{7}}x_{{8}}}{2240}}+
{\frac {y_{{1}}y_{{2}}x_{{3}}x_{{4}}x_{{5}}y_{{6}}x_{{7}}x_{{8}}}{10080}}-
{\frac {y_{{1}}y_{{2}}x_{{3}}y_{{4}}y_{{5}}y_{{6}}y_{{7}}x_{{8}}}{4032}}+
{\frac {y_{{1}}y_{{2}}x_{{3}}x_{{4}}y_{{5}}y_{{6}}y_{{7}}x_{{8}}}{10080}}
\nonumber\\
&&
+{\frac {y_{{1}}y_{{2}}x_{{3}}x_{{4}}x_{{5}}y_{{6}}y_{{7}}x_{{8}}}{10080}}-
{\frac {y_{{1}}y_{{2}}x_{{3}}x_{{4}}x_{{5}}x_{{6}}y_{{7}}x_{{8}}}{4032}}-
{\frac {y_{{1}}y_{{2}}x_{{3}}y_{{4}}x_{{5}}x_{{6}}x_{{7}}x_{{8}}}{4032}}+
{\frac {y_{{1}}x_{{2}}y_{{3}}x_{{4}}x_{{5}}x_{{6}}x_{{7}}x_{{8}}}{10080}}-
{\frac {y_{{1}}x_{{2}}y_{{3}}y_{{4}}x_{{5}}x_{{6}}x_{{7}}x_{{8}}}{4032}}
\nonumber\\
&&
-{\frac {y_{{1}}x_{{2}}x_{{3}}y_{{4}}x_{{5}}x_{{6}}x_{{7}}x_{{8}}}{4032}}+
{\frac {y_{{1}}x_{{2}}y_{{3}}y_{{4}}y_{{5}}x_{{6}}x_{{7}}x_{{8}}}{3024}}+
{\frac {y_{{1}}x_{{2}}x_{{3}}y_{{4}}y_{{5}}x_{{6}}x_{{7}}x_{{8}}}{10080}}-
{\frac {y_{{1}}x_{{2}}x_{{3}}x_{{4}}x_{{5}}y_{{6}}y_{{7}}x_{{8}}}{4032}}+
{\frac {y_{{1}}x_{{2}}x_{{3}}x_{{4}}x_{{5}}x_{{6}}y_{{7}}x_{{8}}}{10080}}
\nonumber\\
&&
+{\frac {y_{{1}}x_{{2}}x_{{3}}x_{{4}}y_{{5}}x_{{6}}x_{{7}}x_{{8}}}{3024}}-
{\frac {y_{{1}}x_{{2}}y_{{3}}y_{{4}}y_{{5}}y_{{6}}x_{{7}}x_{{8}}}{4032}}+
{\frac {y_{{1}}x_{{2}}x_{{3}}y_{{4}}y_{{5}}y_{{6}}x_{{7}}x_{{8}}}{10080}}+
{\frac {y_{{1}}x_{{2}}x_{{3}}x_{{4}}y_{{5}}y_{{6}}x_{{7}}x_{{8}}}{10080}}-
{\frac {y_{{1}}x_{{2}}x_{{3}}x_{{4}}x_{{5}}y_{{6}}x_{{7}}x_{{8}}}{4032}}
\nonumber\\
&&
+{\frac {y_{{1}}x_{{2}}y_{{3}}y_{{4}}y_{{5}}y_{{6}}y_{{7}}x_{{8}}}{10080}}-
{\frac {y_{{1}}x_{{2}}x_{{3}}y_{{4}}y_{{5}}y_{{6}}y_{{7}}x_{{8}}}{4032}}+
{\frac {y_{{1}}x_{{2}}x_{{3}}x_{{4}}y_{{5}}y_{{6}}y_{{7}}x_{{8}}}{3024}}-
{\frac {x_{{1}}x_{{2}}y_{{3}}x_{{4}}x_{{5}}y_{{6}}y_{{7}}y_{{8}}}{10080}}-
{\frac {x_{{1}}x_{{2}}y_{{3}}x_{{4}}x_{{5}}x_{{6}}y_{{7}}y_{{8}}}{10080}}
\nonumber\\
&&
-{\frac {x_{{1}}x_{{2}}y_{{3}}y_{{4}}x_{{5}}y_{{6}}y_{{7}}y_{{8}}}{10080}}-
{\frac {x_{{1}}x_{{2}}x_{{3}}y_{{4}}x_{{5}}x_{{6}}y_{{7}}y_{{8}}}{10080}}-
{\frac {x_{{1}}x_{{2}}y_{{3}}y_{{4}}y_{{5}}x_{{6}}y_{{7}}y_{{8}}}{10080}}-
{\frac {x_{{1}}x_{{2}}y_{{3}}y_{{4}}y_{{5}}x_{{6}}x_{{7}}y_{{8}}}{10080}}-
{\frac {x_{{1}}x_{{2}}x_{{3}}y_{{4}}y_{{5}}x_{{6}}y_{{7}}y_{{8}}}{10080}}
\nonumber\\
&&
-{\frac {x_{{1}}x_{{2}}x_{{3}}y_{{4}}y_{{5}}x_{{6}}x_{{7}}y_{{8}}}{10080}}-
{\frac {x_{{1}}x_{{2}}y_{{3}}y_{{4}}x_{{5}}x_{{6}}x_{{7}}y_{{8}}}{10080}}+
{\frac {x_{{1}}x_{{2}}x_{{3}}x_{{4}}y_{{5}}y_{{6}}x_{{7}}y_{{8}}}{4032}}-
{\frac {x_{{1}}x_{{2}}x_{{3}}x_{{4}}x_{{5}}y_{{6}}x_{{7}}y_{{8}}}{10080}}+
{\frac {x_{{1}}x_{{2}}x_{{3}}x_{{4}}y_{{5}}x_{{6}}y_{{7}}y_{{8}}}{4032}}
\nonumber\\
&&
+{\frac {x_{{1}}x_{{2}}x_{{3}}x_{{4}}y_{{5}}x_{{6}}x_{{7}}y_{{8}}}{4032}}-
{\frac {y_{{1}}y_{{2}}y_{{3}}y_{{4}}y_{{5}}y_{{6}}x_{{7}}x_{{8}}}{60480}}-
{\frac {x_{{1}}y_{{2}}y_{{3}}y_{{4}}y_{{5}}y_{{6}}x_{{7}}y_{{8}}}{10080}}+
{\frac {x_{{1}}x_{{2}}y_{{3}}y_{{4}}y_{{5}}y_{{6}}x_{{7}}y_{{8}}}{4032}}-
{\frac {x_{{1}}x_{{2}}x_{{3}}y_{{4}}y_{{5}}y_{{6}}x_{{7}}y_{{8}}}{3024}}
\nonumber\\
&&
-{\frac {23\,y_{{1}}y_{{2}}y_{{3}}y_{{4}}x_{{5}}x_{{6}}x_{{7}}x_{{8}}}{120960}}-
{\frac {x_{{1}}y_{{2}}y_{{3}}y_{{4}}x_{{5}}y_{{6}}y_{{7}}y_{{8}}}{3024}}-{
\frac {x_{{1}}y_{{2}}y_{{3}}y_{{4}}x_{{5}}x_{{6}}x_{{7}}y_{{8}}}{3024}}-
{\frac {x_{{1}}x_{{2}}y_{{3}}y_{{4}}x_{{5}}x_{{6}}y_{{7}}y_{{8}}}{2240}}-
{\frac {x_{{1}}x_{{2}}x_{{3}}y_{{4}}x_{{5}}y_{{6}}y_{{7}}y_{{8}}}{3024}}
\nonumber\\
&&
-{\frac {x_{{1}}x_{{2}}x_{{3}}y_{{4}}x_{{5}}x_{{6}}x_{{7}}y_{{8}}}{3024}}+
{\frac {y_{{1}}y_{{2}}y_{{3}}y_{{4}}y_{{5}}x_{{6}}x_{{7}}x_{{8}}}{10080}}+
{\frac {x_{{1}}y_{{2}}y_{{3}}y_{{4}}y_{{5}}x_{{6}}y_{{7}}y_{{8}}}{4032}}+
{\frac {x_{{1}}y_{{2}}y_{{3}}y_{{4}}y_{{5}}x_{{6}}x_{{7}}y_{{8}}}{4032}}+
{\frac {y_{{1}}y_{{2}}y_{{3}}x_{{4}}x_{{5}}x_{{6}}x_{{7}}x_{{8}}}{10080}}
\nonumber\\
&&
+{\frac {x_{{1}}y_{{2}}y_{{3}}x_{{4}}y_{{5}}y_{{6}}y_{{7}}y_{{8}}}{4032}}+
{\frac {x_{{1}}y_{{2}}y_{{3}}x_{{4}}x_{{5}}x_{{6}}x_{{7}}y_{{8}}}{4032}}+
{\frac {x_{{1}}x_{{2}}y_{{3}}x_{{4}}y_{{5}}y_{{6}}y_{{7}}y_{{8}}}{4032}}+
{\frac {x_{{1}}x_{{2}}y_{{3}}x_{{4}}x_{{5}}x_{{6}}x_{{7}}y_{{8}}}{4032}}-
{\frac {y_{{1}}y_{{2}}x_{{3}}x_{{4}}x_{{5}}x_{{6}}x_{{7}}x_{{8}}}{60480}}
\nonumber\\
&&
-{\frac {x_{{1}}y_{{2}}x_{{3}}y_{{4}}y_{{5}}y_{{6}}y_{{7}}y_{{8}}}{10080}}+
{\frac {x_{{1}}y_{{2}}x_{{3}}x_{{4}}y_{{5}}y_{{6}}y_{{7}}y_{{8}}}{4032}}-
{\frac {x_{{1}}y_{{2}}x_{{3}}x_{{4}}x_{{5}}y_{{6}}y_{{7}}y_{{8}}}{3024}}+
{\frac {x_{{1}}y_{{2}}x_{{3}}x_{{4}}x_{{5}}x_{{6}}y_{{7}}y_{{8}}}{4032}}-
{\frac {x_{{1}}y_{{2}}x_{{3}}x_{{4}}x_{{5}}x_{{6}}x_{{7}}y_{{8}}}{10080}}
\nonumber\\
&&
-{\frac {x_{{1}}y_{{2}}y_{{3}}y_{{4}}x_{{5}}x_{{6}}y_{{7}}y_{{8}}}{10080}}
-{\frac {x_{{1}}y_{{2}}y_{{3}}x_{{4}}x_{{5}}y_{{6}}y_{{7}}y_{{8}}}{10080}}-
{\frac {x_{{1}}y_{{2}}y_{{3}}x_{{4}}x_{{5}}x_{{6}}y_{{7}}y_{{8}}}{10080}}
+{\frac {x_{{1}}x_{{2}}y_{{3}}y_{{4}}y_{{5}}y_{{6}}y_{{7}}y_{{8}}}{60480}}
-{\frac {x_{{1}}x_{{2}}x_{{3}}y_{{4}}y_{{5}}y_{{6}}y_{{7}}y_{{8}}}{10080}}
\nonumber\\
&&
+{\frac {23\,x_{{1}}x_{{2}}x_{{3}}x_{{4}}y_{{5}}y_{{6}}y_{{7}}y_{{8}}}{120960}}
-{\frac {x_{{1}}x_{{2}}x_{{3}}x_{{4}}x_{{5}}y_{{6}}y_{{7}}y_{{8}}}{10080}}+
{\frac {x_{{1}}x_{{2}}x_{{3}}x_{{4}}x_{{5}}x_{{6}}y_{{7}}y_{{8}}}{60480}}+
{\frac {y_{{1}}x_{{2}}y_{{3}}x_{{4}}y_{{5}}x_{{6}}y_{{7}}x_{{8}}}{280}}.
\end{eqnarray}
There are 62 occurrences of each $x_i$ above, and 62 occurrences of each $y_i$.
(Note that $62+62=124 = \hbox{number of terms}$.)
Furthermore the $x_{{1}}x_{{2}}x_{{3}}x_{{4}}y_{{5}}y_{{6}}y_{{7}}y_{{8}}$ and $y_{{1}}y_{{2}}y_{{3}}y_{{4}}x_{{5}}x_{{6}}x_{{7}}x_{{8}}$ terms are the only two terms above with a nontrivial integer in the \emph{numerator}; they are the first such terms in the general Baker--Campbell--Hausdorff expansion. 

\clearpage

Converting $x_i\to X$ and $y_i \to Y$ so as to get a representation in terms of words in the $\{X,Y\}$ alphabet we obtain:
 {\small
\begin{eqnarray}
{\large z_8(X,Y) }&=&
{\frac {Y^3XYXYX}{1260}}-
{\frac {XYXYXYXY}{280}}-
{\frac {Y^2XY^2XYX}{1680}}-
{\frac {Y^2X^2YXYX}{1680}}-
{\frac {Y^2XYXYX^2}{1680}}
\nonumber\\
&&
-{\frac {Y^2XYXY^2X}{1680}}-
{\frac {Y^2XYX^2YX}{1680}}-
{\frac {YXY^2XYX^2}{1680}}-
{\frac {YXY^2XY^2X}{1680}}-
{\frac {YXY^2X^2YX}{1680}}
\nonumber\\
&&
-{\frac {YXYX^2YX^2}{1680}}+
{\frac {YXYXY^3X}{1260}}-
{\frac {YXYX^2Y^2X}{1680}}+
{\frac {YXYX^3YX}{1260}}+
{\frac {YXYXYX^3}{1260}}
\nonumber\\
&&
-{\frac {YXYXY^2X^2}{1680}}+
{\frac {YXY^3XYX}{1260}}-
{\frac {YX^2Y^2XYX}{1680}}+
{\frac {YX^3YXYX}{1260}}-
{\frac {YX^2YXYX^2}{1680}}
\nonumber\\
&&
-{\frac {YX^2YXY^2X}{1680}}-
{\frac {YX^2YX^2YX}{1680}}+
{\frac {Y^5XYX}{10080}}+
{\frac {X^2Y^2XYXY}{1680}}-
{\frac {X^3YXYXY}{1260}}
\nonumber\\
&&
-{\frac {Y^4XYX^2}{4032}}-
{\frac {Y^4XY^2X}{4032}}-
{\frac {Y^4X^2YX}{4032}}-
{\frac {XY^3XYXY}{1260}}+
{\frac {X^2YX^2YXY}{1680}}
\nonumber\\
&&
+{\frac {XY^2XY^2XY}{1680}}+
{\frac {XY^2X^2YXY}{1680}}+
{\frac {X^2YXYXY^2}{1680}}+
{\frac {X^2YXYX^2Y}{1680}}+
{\frac {X^2YXY^2XY}{1680}}
\nonumber\\
&&
+{\frac {Y^3XYX^3}{3024}}+
{\frac {Y^3XY^2X^2}{10080}}+
{\frac {Y^3X^2YX^2}{10080}}+
{\frac {XY^2XYXY^2}{1680}}+
{\frac {XY^2XYX^2Y}{1680}}
\nonumber\\
&&
+{\frac {Y^3XY^3X}{3024}}+
{\frac {Y^3X^2Y^2X}{10080}}+
{\frac {Y^3X^3YX}{3024}}-
{\frac {XYXYXY^3}{1260}}+
{\frac {XYXYX^2Y^2}{1680}}
\nonumber\\
&&
-{\frac {XYXYX^3Y}{1260}}+
{\frac {XYXY^2XY^2}{1680}}+
{\frac {XYXY^2X^2Y}{1680}}+
{\frac {XYX^2YXY^2}{1680}}+
{\frac {XYX^2YX^2Y}{1680}}
\nonumber\\
&&
-{\frac {XYXY^3XY}{1260}}+
{\frac {XYX^2Y^2XY}{1680}}-
{\frac {XYX^3YXY}{1260}}+
{\frac {Y^2XY^2X^3}{10080}}+
{\frac {Y^2X^2YX^3}{10080}}
\nonumber\\
&&
+{\frac {Y^2XY^3X^2}{10080}}+
{\frac {Y^2X^2Y^2X^2}{2240}}+
{\frac {Y^2X^3YX^2}{10080}}-
{\frac {Y^2XY^4X}{4032}}+
{\frac {Y^2X^2Y^3X}{10080}}
\nonumber\\
&&
+{\frac {Y^2X^3Y^2X}{10080}}-
{\frac {Y^2X^4YX}{4032}}-
{\frac {Y^2XYX^4}{4032}}+
{\frac {YXYX^5}{10080}}-
{\frac {YXY^2X^4}{4032}}
\nonumber\\
&&
-{\frac {YX^2YX^4}{4032}}+
{\frac {YXY^3X^3}{3024}}+
{\frac {YX^2Y^2X^3}{10080}}-
{\frac {YX^4Y^2X}{4032}}+
{\frac {YX^5YX}{10080}}
\nonumber\\
&&
+{\frac {YX^3YX^3}{3024}}-
{\frac {YXY^4X^2}{4032}}+
{\frac {YX^2Y^3X^2}{10080}}+
{\frac {YX^3Y^2X^2}{10080}}-
{\frac {YX^4YX^2}{4032}}
\nonumber\\
&&
+{\frac {YXY^5X}{10080}}-
{\frac {YX^2Y^4X}{4032}}+
{\frac {YX^3Y^3X}{3024}}-
{\frac {X^2YX^2Y^3}{10080}}-
{\frac {X^2YX^3Y^2}{10080}}
\nonumber\\
&&
-{\frac {X^2Y^2XY^3}{10080}}-
{\frac {X^3YX^2Y^2}{10080}}-
{\frac {X^2Y^3XY^2}{10080}}-
{\frac {X^2Y^3X^2Y}{10080}}-
{\frac {X^3Y^2XY^2}{10080}}
\nonumber\\
&&
-{\frac {X^3Y^2X^2Y}{10080}}-
{\frac {X^2Y^2X^3Y}{10080}}+
{\frac {X^4Y^2XY}{4032}}-
{\frac {X^5YXY}{10080}}+
{\frac {X^4YXY^2}{4032}}
\nonumber\\
&&
+{\frac {X^4YX^2Y}{4032}}-
{\frac {Y^6X^2}{60480}}-
{\frac {XY^5XY}{10080}}+
{\frac {X^2Y^4XY}{4032}}-
{\frac {X^3Y^3XY}{3024}}
\nonumber\\
&&
-{\frac {23\,Y^4X^4}{120960}}-
{\frac {XY^3XY^3}{3024}}-{
\frac {XY^3X^3Y}{3024}}-
{\frac {X^2Y^2X^2Y^2}{2240}}-
{\frac {X^3YXY^3}{3024}}
\nonumber\\
&&
-{\frac {X^3YX^3Y}{3024}}+
{\frac {Y^5X^3}{10080}}+
{\frac {XY^4XY^2}{4032}}+
{\frac {XY^4X^2Y}{4032}}+
{\frac {Y^3X^5}{10080}}
\nonumber\\
&&
+{\frac {XY^2XY^4}{4032}}+
{\frac {XY^2X^4Y}{4032}}+
{\frac {X^2YXY^4}{4032}}+
{\frac {X^2YX^4Y}{4032}}-
{\frac {Y^2X^6}{60480}}
\nonumber\\
&&
-{\frac {XYXY^5}{10080}}+
{\frac {XYX^2Y^4}{4032}}-
{\frac {XYX^3Y^3}{3024}}+
{\frac {XYX^4Y^2}{4032}}-
{\frac {XYX^5Y}{10080}}
\nonumber\\
&&
-{\frac {XY^3X^2Y^2}{10080}}
-{\frac {XY^2X^2Y^3}{10080}}-
{\frac {XY^2X^3Y^2}{10080}}
+{\frac {X^2Y^6}{60480}}
-{\frac {X^3Y^5}{10080}}
\nonumber\\
&&
+{\frac {23\,X^4Y^4}{120960}}
-{\frac {X^5Y^3}{10080}}+
{\frac {X^6Y^2}{60480}}+
{\frac {YXYXYXYX}{280}}
\end{eqnarray}
}
\enlargethispage{20pt}\!
There are 2 occurrences of $X^6$ above,
6 occurrences of $X^5$,
14 occurrences of $X^4$,
32 occurrences of $X^3$,
72 occurrences of $X^2$,
and 158 occurrences of $X^1$.
Similarly for $Y$.
(Note that $62\times 8 = 2\times 6 + 6\times5+14\times4+32\times3+72\times2+158\times 1$.)
Converting this to commutators, (in any form, either the Dynkin form or using other commutator bases),  is impractical without significant additional computer-aided computation. See for instance reference~\cite{Casas-Murua}, and related online tables~\cite{Murua-tables}. 

\clearpage


\end{document}